\documentclass[11pt, journal]{IEEEtran}
\usepackage{algorithmicx}
\usepackage{amsmath}
\usepackage{amsfonts}
\usepackage{stfloats}

\usepackage{hyperref}
 
\usepackage{enumitem}
\usepackage{algpseudocode}
\usepackage{multirow}
\usepackage{url}
\usepackage{epsfig}
\usepackage{multicol,lipsum,xparse}
\usepackage{subcaption}
\usepackage{amsmath}
\usepackage{stmaryrd}
\usepackage{color}
\usepackage{todonotes}
\usepackage[utf8]{inputenc}
\usepackage{tikz}
\usetikzlibrary{decorations.pathreplacing}
\usepackage{graphicx}
\usepackage{colortbl}
\usepackage{tabulary}
\usepackage{etoolbox}
\usepackage[linesnumbered,ruled,vlined]{algorithm2e}
\SetAlgoNlRelativeSize{-1}
\SetAlgoNlRelativeSize{-2}
\SetAlgoNlRelativeSize{-3}

\usepackage{tikz}

\title{DAG-aware Synthesis Orchestration}
\vspace{4mm}

\author{Yingjie Li, Mingju Liu, \textit{IEEE Student Member}, Haoxing Ren, Alan Mishchenko, \textit{IEEE Senior Member}, Cunxi Yu, \textit{IEEE Member}}
\newbool{journalBasedAuthors}
\ifCLASSOPTIONpeerreview \setbool{journalBasedAuthors}{true}
\else \ifCLASSOPTIONjournal \setbool{journalBasedAuthors}{true}
\else \setbool{journalBasedAuthors}{false}
\fi
\fi

\ifbool{journalBasedAuthors}
{

    \author{Yingjie~Li,
            Mingju~Liu,
            Haoxing~Ren,
            Alan~Mishchenko, 
            Cunxi~Yu
    \thanks{Y. Li, M. Liu and C. Yu are with the Department of Electrical and Computer Engineering, University of Maryland,  College Park, US (e-mails: yingjieli@umd.edu, mliu9867@umd.edu cunxiyu@umd.edu). A. Mishchenko is with the Department of Electrical Engineering and Computer Sciences, University of California, Berkeley, US (e-mail: alanmi@berkeley.edu). H. Ren is with Nvidia Research, Austin, Texas, US (e-mail: haoxingr@nvidia.com).}
    \thanks{This work is funded by National Science Foundation (NSF) NSF-2008144 and NSF CAREER award NSF-2047176.}
    \ifCLASSOPTIONjournal
        \thanks{Digital Object Identifier 10.1109/TCAD.XXXXXXX}
    \fi
    }
}

\begin{document}
 \newcommand{\pgftextcircled}[1]{
    \setbox0=\hbox{#1}%
    \dimen0\wd0%
    \divide\dimen0 by 2%
    \begin{tikzpicture}[baseline=(a.base)]%
        \useasboundingbox (-\the\dimen0,0pt) rectangle (\the\dimen0,1pt);
        \node[circle,draw,outer sep=0pt,inner sep=0.1ex] (a) {#1};
    \end{tikzpicture}
}

\maketitle

\begin{abstract}

Modern logic synthesis techniques use multi-level technology-independent representations like And-Inverter-Graphs (AIGs) for digital logic. This involves structural rewriting, resubstitution, and refactoring based on directed-acyclic-graph (DAGs) traversal. {Existing DAG-aware logic synthesis algorithms are designed to perform one specific optimization during a single DAG traversal}. However, we empirically identify and demonstrate that these algorithms are limited in quality-of-results due to the solely considered optimization operation in the design concept. {This work proposes Synthesis \textit{Orchestration}, which is a fine-grained node-level optimization implying multiple optimizations during the single traversal of the graph.} %Thus, \textit{orchestration} method explores more optimization opportunities and results in better performance. 
Our experimental results are comprehensively conducted on all 104 designs collected from ISCAS'85/89/99, VTR, and EPFL benchmark suites. The orchestration algorithms consistently outperform existing optimizations, \textit{rewriting}, \textit{resubstitution}, \textit{refactoring}, leading to an average of 4\% more node reduction with reasonable runtime cost for the single optimization. Moreover, we evaluate the orchestration algorithm in the sequential optimization, and as a plug-in algorithm in \textit{resyn} and \textit{resyn3} flows in ABC, which demonstrate consistent logic minimization improvements (1\%, 4.7\% and 11.5\% more node reduction on average). Finally, we integrate the orchestration into OpenROAD for end-to-end performance evaluations. Our results demonstrate the advantages of the orchestration optimization techniques, even after technology mapping and post-routing in the design flow.

\end{abstract}

\section{Introduction}
\label{sec:introduction}

Logic optimization plays a critical role in design automation flows for digital systems, significantly impacting area, timing closure, and power optimizations~\cite{bjesse2004dag, haaswijk2018deep, brayton1990multilevel, amaru2015majority, mishchenko2007abc, yu2020flowtune, yu2018developing}, as well as influencing new trends in neural network optimizations~\cite{rai2021logic, huang2022quantized}. The goal of logic optimization is to achieve higher performance, reduced area, and lower power consumption, all while maintaining the original functionality of the circuit.

Modern digital designs are complex and feature with millions of logic gates, coupled with an extensive exploration space. This complexity underscores the importance of efficient, technology-independent optimizations for design area and delay at the logic level. Key methodologies in modern logic optimization techniques are conducted on multi-level, technology-independent representations, such as And-Inverter-Graphs (AIGs)~\cite{mishchenko2006dag, mishchenko2005fraigs, yu2016dag} and Majority-Inverter-Graphs (MIGs)~\cite{amaru2015majority, soeken2017exact}, for digital logic. Additionally, XOR-rich representations are crucial for emerging technologies, as seen in XOR-And-Graphs~\cite{ccalik2019multiplicative} and XOR-Majority-Graphs~\cite{haaswijk2017novel}.

\begin{figure*}
    \centering
    \begin{subfigure}[b]{0.23\textwidth}
    \centering
        \includegraphics[width=1\textwidth]{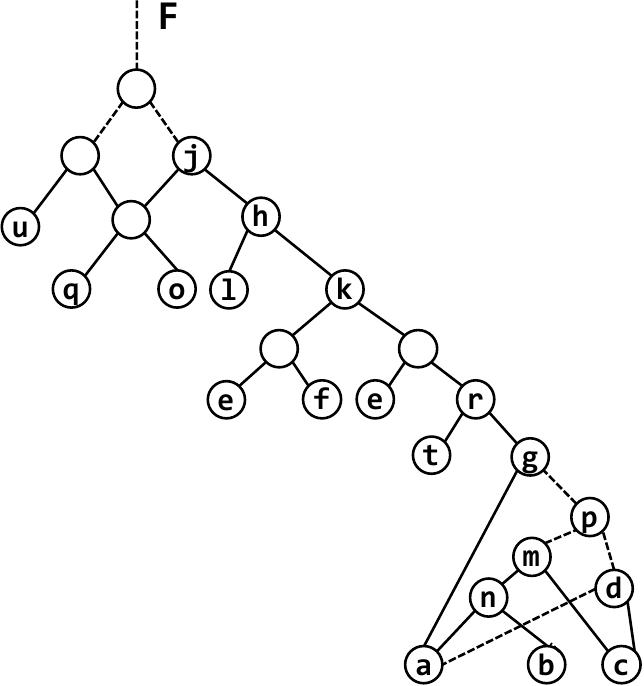}
        \caption{original AIG}
        \label{fig:aig_ori}
    \end{subfigure}
    \hfill
    \begin{subfigure}[b]{0.23\textwidth}
    \centering
        \includegraphics[width=1\textwidth]{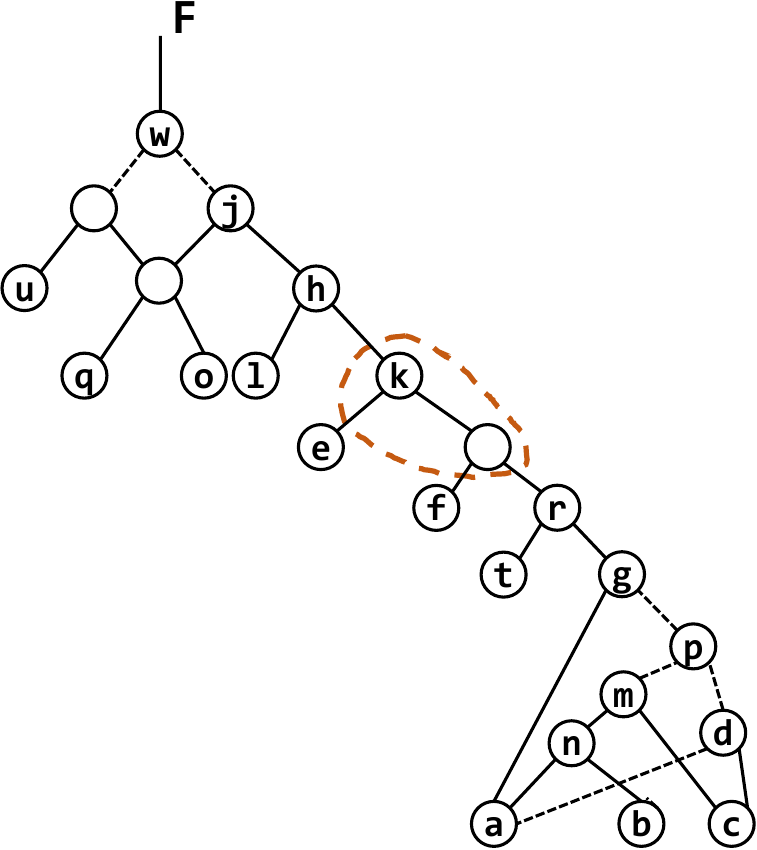}
        \caption{AIG with \texttt{rw}}
        \label{fig:aig_rw}
    \end{subfigure}
    \hfill
    \begin{subfigure}[b]{0.23\textwidth}
    \centering
\includegraphics[width=1\textwidth]{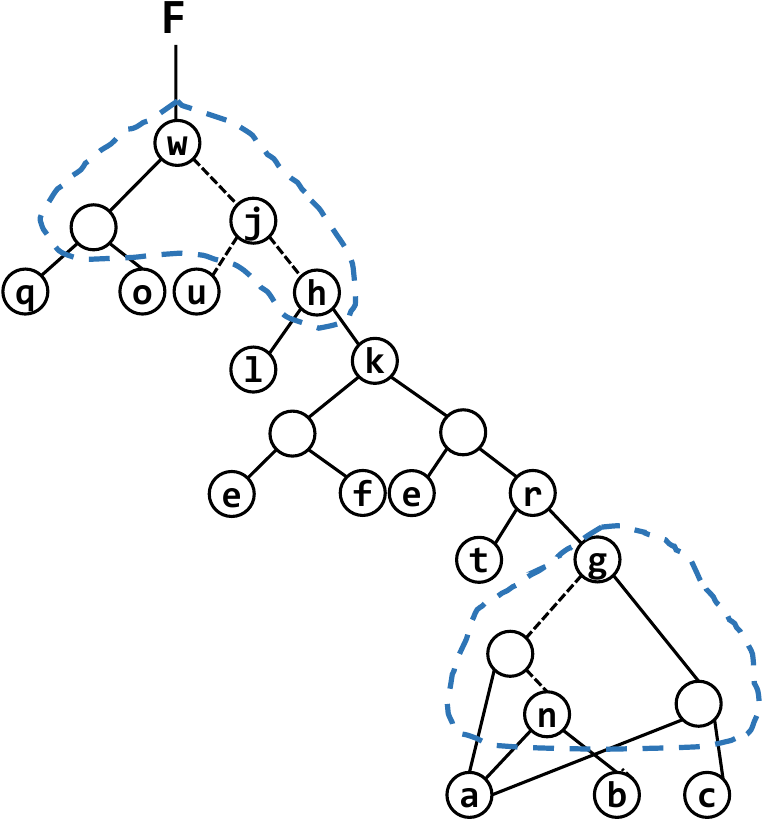}
        \caption{AIG with \texttt{rf}}
        \label{fig:aig_rf}
    \end{subfigure}
    \hfill
    \begin{subfigure}[b]{0.23\textwidth}
    \centering
\includegraphics[width=1\textwidth]{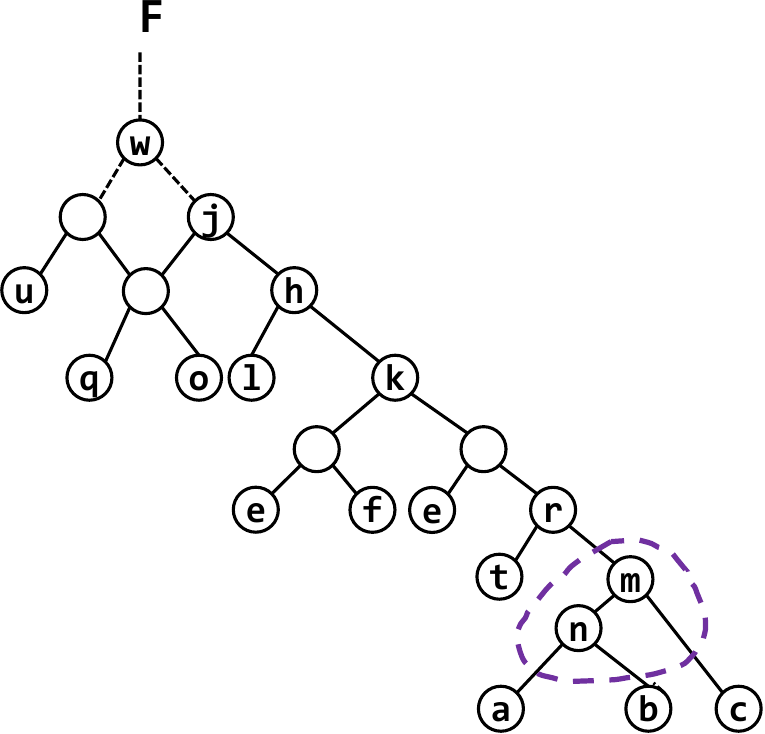}
        \caption{AIG with \texttt{rs}}
        \label{fig:aig_rs}
    \end{subfigure}
    
    \caption{{The optimized graph produced by stand-alone optimization operations: (a) original AIG, graph size is 25; (b) optimized AIG with stand-alone \texttt{rw}, graph size is 23; (c) optimized AIG with stand-alone \texttt{rf}, graph size is 23; (d) optimized AIG with stand-alone \texttt{rs}, graph size is 22.}
    }
    \label{fig:aig_alone}
\end{figure*}

A framework for logic synthesis, {ABC}~\cite{mishchenko2007abc}, introduces multiple state-of-the-art (SOTA) Directed-Acyclic-Graphs (DAGs) aware Boolean optimization algorithms. These include structural rewriting (command \textit{rewrite} in ABC)~\cite{mishchenko2006dag, riener2022boolean, haaswijk2018sat}, resubstitution (command \textit{resub} in ABC)~\cite{brayton2006scalable}, and refactoring (command \textit{refactor} in ABC)~\cite{mishchenko2006dag}, all of which are based on the AIG data structure. {During the existing logic optimization process, the algorithm considers a single specific optimization method and applies the optimization based on a single criterion~\cite{mishchenko2006dag}. }%For instance, during the single AIG traversal in the \textit{rewrite} algorithm, each node is evaluated against the optimization criterion of structural rewriting, and optimization is applied accordingly. Even though some logic cuts might be optimized using other methods, such as \textit{refactor} and \textit{resub}, these optimization opportunities are overlooked during the singular \textit{rewrite} optimization process.} %Therefore, existing optimization methods that consider only one specific optimization within a single traversal of the AIG are referred to as "stand-alone" optimizations in this context.}
Our empirical studies further reveal critical limitations inherent in the mainstream stand-alone concept of logic optimization, particularly in missing significant optimization opportunities. These opportunities are often overlooked due to a consistent tendency of becoming stuck in "bad" local minima. In other words, the optimization of a node, when various applicable optimization opportunities are present, is constrained by the limitations inherent in the current stand-alone optimization concept. For instance, as depicted in Figure \ref{fig:aig_alone}, although node $g$ is suitable for both \textit{refactoring} and \textit{resubstitution}, it misses potential optimization opportunities when subjected solely to \textit{rewriting}. 

{In this work, we propose a novel logic synthesis development concept, \textbf{DAG-aware Synthesis \textit{Orchestration}}, that maximizes optimizations through Boolean transformations by orchestrating multiple optimization operations in the single traversal of the logic graph.} Specifically, we implement the synthesis orchestration approach based on AIGs by orchestrating \textit{rewrite}, \textit{refactor}, and \textit{resub} implemented in ABC~\cite{mishchenko2007abc} in the single optimization command \textit{orchestration}. The orchestration algorithm is orthogonal to other DAG-aware synthesis algorithms, which can be applied to Boolean networks independently and/or iteratively. Our results demonstrate that applying orchestration in DAG-aware synthesis can significantly improve logic optimization compared to the existing optimization methods. We anticipate that the concept of logic synthesis orchestration can be effectively extended to other data structures, such as Majority-Inverter Graphs (MIGs)~\cite{amaru2015majority}.
%For each AIG node, \textit{orchestration} provides all valid operations to select from and apply the operation to update the graph, as illustrated in Figure \ref{fig:aig} and Figure \ref{fig:orch_flow}. With our method, the optimization space within the single traversal of the AIG is expanded and more optimization opportunities are explored at the same runtime scale. 

The main contributions of the work are summarized as follows:
\begin{itemize} 
    \item Our comprehensive analysis and examples (Figures \ref{fig:aig_alone} and \ref{fig:ana}) highlight significant optimization losses in current logic optimization implementations. %Comprehensive node optimization opportunity analysis (Figures \ref{fig:aig_alone} and \ref{fig:ana}) and motivating examples are provided, which demonstrate significant optimization opportunity loss in existing SOTA implementations.
    
    \item We propose two DAG-aware synthesis orchestration algorithms, \textit{Priority-ordered orchestration} and \textit{Local-greedy orchestration} to define the criteria for orchestrating \textit{rewrite}, \textit{refactor}, and \textit{resub} in AIG optimizations (Section \ref{sec:approach}).
    
    \item We provide the performance evaluations and runtime analysis on 104 designs from five benchmark suites (ISCAS'85/89 \cite{brglez1989combinational}, ITC/ISCAS'99 \cite{davidson1999itc}, VTR \cite{murray2020vtr}, and EPFL benchmarks \cite{soeken2018epfl}), which shows our {orchestration} technique achieves an average of 4.2\% more AIG reductions compared to existing logic optimization algorithms in ABC (Section \ref{sec:single_result} and \ref{sec:runtime_result}). %We provide comprehensive evaluation and explorations of proposed \texttt{orchestration} on 104 designs collected from five benchmark suites, including ISCAS'85/89 \cite{brglez1989combinational}, ITC/ISCAS'99 \cite{davidson1999itc}, VTR \cite{murray2020vtr}, and EPFL benchmarks \cite{soeken2018epfl}, with an average of 4.2\% more AIG reductions compared to SOTA stand-alone optimizations. %, and provides $11.7\%$ optimization in graph size compared to \textit{resyn}, and $2.24\times$ compared to \textit{resyn3} .
    \item We provide the evaluations of sequential optimizations with {orchestration} algorithms, where the {orchestration} techniques show its performance advantage of 4.7\% for \texttt{resyn} and 11.5\% for \texttt{resyn3} (Section \ref{sec:iterative_result}). 
    \item We further integrate orchestrated logic optimizations into OpenROAD \cite{ajayi2019openroad} for end-to-end design evaluations, demonstrating consistent AIG minimization and area improvements for post-technology mapping and routing (Section \ref{sec:eval_openroad}). %We further integrate the orchestrated logic optimization into OpenROAD \cite{ajayi2019openroad} to evaluate the \texttt{orchestration} performance w.r.t the end-to-end chip design flow, demonstrating consistent AIG minimization and area improvements for post technology mapping and post routing.
    \item Our approach is available in ABC \cite{mishchenko2007abc} through a new command, \textit{orchestration}.
\end{itemize}

\section{Preliminary}

\subsection{Boolean Networks and AIGs}
\label{sec:background_BN}
A Boolean network is a directed acyclic graph (DAG) denoted as $G=(V, E)$ with nodes $V$ representing logic gates (Boolean functions) and edges $E$ representing the wire connection between gates. The input of a node is called its \textit{fanin}, and the output of the node is called its \textit{fanout}. The node $v \in V$ without incoming edges, i.e., no \textit{fanins}, is the \textit{primary input} (PI) to the graph, and the nodes without outgoing edges, i.e., no \textit{fanouts}, are \textit{primary outputs} (POs) to the graph. The nodes with incoming edges implement Boolean functions. The level of a node $v$ is defined by the number of nodes on the longest structural path from any PI to the node inclusively, and the level of a node $v$ is noted as $level(v)$.

%\subsection{And-Inverter Graph (AIG)}
%\label{sec:background_AIG}
\textit{And-Inverter Graph} (AIG) is one of the typical types of DAGs used for logic manipulations, where the nodes in AIGs are all two-inputs AND gates, and the edges represent whether the inverters are implemented. An arbitrary Boolean network can be transformed into an AIG by factoring the SOPs of the nodes, and the AND gates and OR gates in SOPs are converted to two-inputs AND gates and inverters with DeMorgan's rule. There are two primary metrics for evaluation of an AIG, i.e., \textit{size}, which is the number of nodes (AND gates) in the graph, and \textit{depth}, which is the number of nodes on the longest path from PI to PO (the largest level) in the graph. 
{A \textit{cut} $C$ of node $v$ includes a set of nodes of the network. The leaf nodes included in the \textit{cut} of node $v$ are called \textit{leaves}, such that each path from a PI to node $v$ passes through at least one leaf. The node $v$ is called the \textit{root} node of the \textit{cut} C. The cut size is the number of its leaves. A cut is $K$-feasible if the number of leaves in the cut does not exceed $K$.}
%called leaves, such that each path from a PI to n passes through at least one leaf. Node n is called root of cut C. The cut size is the number of its leaves. A trivial cut of the node is the cut composed of the node itself. A cut is K-feasible if the number of nodes in the cut does not exceed K. A cut is said to be dominated if there is another cut of the same node, which is contained, set-theoretically, in the given cut. The volume of a cut is the total number of nodes encountered on all paths between node n and the cut leaves.
The logic optimization of Boolean networks can be conducted with the AIGs efficiently~\cite{hosny2020drills, yu2018developing} based on the Boolean algebra enabled logic transformations.

%The AIG-based optimization process in the single traversal of the graph usually involves two steps: \textbf{(1)} \textit{transformability check} -- check the transformability w.r.t the optimization operation for the current node; \textbf{(2)} \textit{graph updates} {(\texttt{Dec\_GraphUpdateNetwork} in ABC)} -- apply the applicable optimization operation at the node to realize the transformation and update the graph for the next unseen node. %\cy{I think we could link these two steps with pseduo code syntax that is used in our Algs 1 and 2, e.g., \texttt{Dec\_GraphUpdateNetwork}}

\subsection{DAG-Aware Logic Synthesis}
\label{sec:background_DAG_syn}

To minimize logic complexity and size, subsequently leading to enhanced performance, DAG-aware logic optimization approaches leverage Boolean algebra at direct-acyclic-graph (DAG) logic representations, aiming to minimize area, power, delay, etc., while preserving the original functionality of the circuit.

This is achieved through the application of various technology-independent optimization techniques and algorithms, such as node rewriting, structural hashing, and refactoring. In this work, we focus specifically on exploring DAG-aware logic synthesis using And-Inverter Graphs (AIGs) representations. 
{The AIG-based optimization process, during a single traversal of the logic graph, typically involves two steps: \textbf{(1)} \textit{transformability check} -- checking the transformability of the optimization operation for the logic cut in relation to the current node; \textbf{(2)} \textit{graph updates} -- if the optimization is applicable, the optimization operation is applied at the node to realize the transformation of the logic cut and subsequently update the graph.}

\noindent
\textbf{Rewriting~\cite{mishchenko2006dag}}, denoted as \texttt{rw}, {is a fast greedy algorithm for logic optimization.} {It iteratively selects an AIG logic cut with the current node as the root node and replaces the selected subgraph with the same functional pre-computed subgraph (NPN-equivalent) of a smaller size to realize the graph size optimization. In the default settings in ABC, the target logic cuts for each node are 4-feasible cuts. For AIG rewriting, all 4-feasible cuts of the nodes are pre-computed using the fast cut enumeration procedure. In each iteration, the Boolean function for the current logic cut is computed and its NPN-class is determined by hash-table lookup.
After trying all available subgraphs, the one that leads to the largest improvement at a node is used.}
For instance, Figure \ref{fig:aig_rw} illustrates the optimization of the original graph shown in Figure \ref{fig:aig_ori} using \texttt{rw}. The algorithm traverses each node in topological order, checking the transformability of its cut with \textit{rewriting}. In Figure \ref{fig:aig_rw}, node $k=efr$ is optimized using \texttt{rw}, resulting in a reduction of 2 nodes for the logic optimization.

\noindent
\textbf{Refactoring~\cite{mishchenko2006dag}}, denoted as \texttt{rf}, is a variation of the AIG \textit{rewriting} using a heuristic algorithm to produce a larger cut for each AIG node. Refactoring optimizes AIGs by replacing the current AIG structure with a factored form of the cut function.
For example, Figure \ref{fig:aig_rf} illustrates the optimization of the original AIG with \texttt{rf}. {Node $g$ is optimized to the factored form of $g=ac(\overline{n} + a)$ and the node $w$ is optimized to $w=qo({u} + {h})$}. As a result, the optimized graph with \texttt{rf} has a graph size of $23$ with $2$ nodes reduction.

\noindent
\textbf{Resubstitution~\cite{brayton2006scalable}}, denoted as \texttt{rs}, optimizes the AIG by replacing the function of a node with functions of other existing nodes, referred as {divisors}, within the graph. This approach aims to eliminate redundant nodes unnecessary for expressing the function of the current node. {In \textit{resubstitution}, cuts containing no more than 12-16 leaves are considered, and the optimization is performed using explicitly computed truth tables and exhaustive simulation. During \textit{resubstitution}, the introduction of new nodes may occur to complete the functionality in the AIG, which is a process known as $k$-resubstitution (where $k$ represents the number of newly introduced nodes) and $k$ should not exceed the number of nodes saved by the optimization. In the default settings of ABC, $k$-resubstitution is checked for $k=\{0, 1, 2, 3\}$, and the number of divisors in each cut is limited to $150$.}
For example, in Figure \ref{fig:aig_ori}, the node $g=a\overline{p}$, with $p=\overline{m}\overline{d}$, ${d}=\overline{a}{c}$, and $m=abc$, implies $g=abc$. This condition allows for resubstitution with node $m$, leading to the removal of nodes $g, p,$ and $d$ from the graph, as depicted in Figure \ref{fig:aig_rs}. Consequently, \texttt{rs} optimizes the original AIG by reducing the graph size through the removal of $3$ nodes.

\noindent
{\textbf{Definition 1: \textit{Stand-alone} Logic Optimization}: Stand-alone logic optimization refers to the process of optimizing the logic graph using a single pre-set optimization criterion during the single traversal of the entire graph.}
{\textbf{Example 1:} The existing optimizations, such as structural \textit{rewriting}, \textit{refactoring}, and \textit{resubstitution}, are stand-alone optimizations as they only assess the transformability with respect to a single pre-set operation and update the graph based on the corresponding optimization criterion.
}

\section{Approach}
\label{sec:approach}

\begin{figure*}
    \centering
    \begin{subfigure}[b]{0.19\textwidth}
        \centering
        \includegraphics[width=1\textwidth]{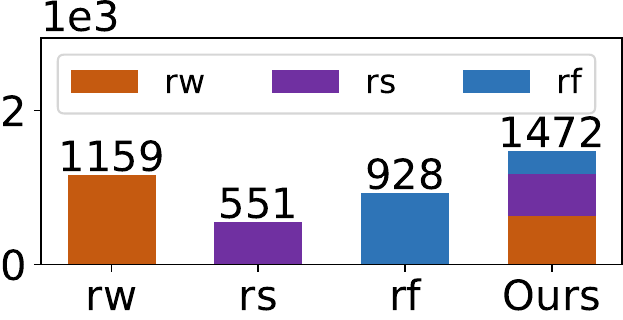}
        \caption{s38584}
    \end{subfigure}
    \hfill
    \begin{subfigure}[b]{0.2\textwidth}
        \centering
        \includegraphics[width=1\textwidth]{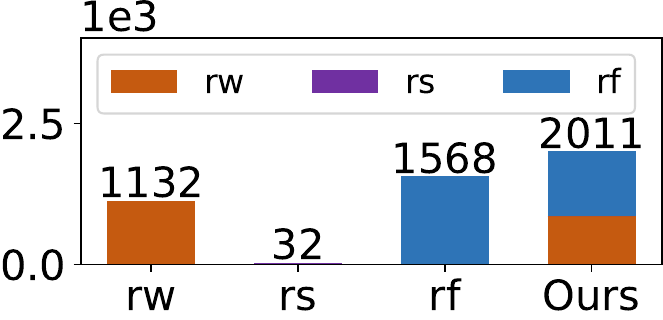}
        \caption{s35932}
    \end{subfigure}
    \hfill
    \begin{subfigure}[b]{0.19\textwidth}
        \centering
        \includegraphics[width=1\textwidth]{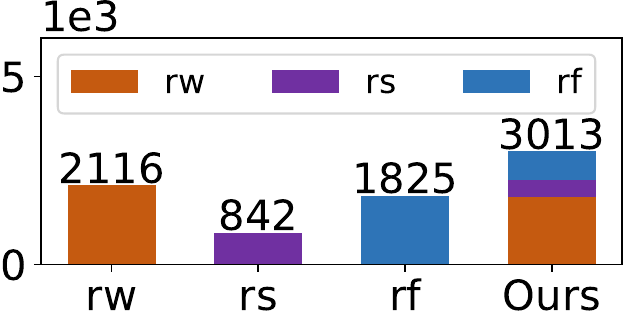}
        \caption{b17\_1}
    \end{subfigure}
    \hfill
    \begin{subfigure}[b]{0.19\textwidth}
        \centering
        \includegraphics[width=1\textwidth]{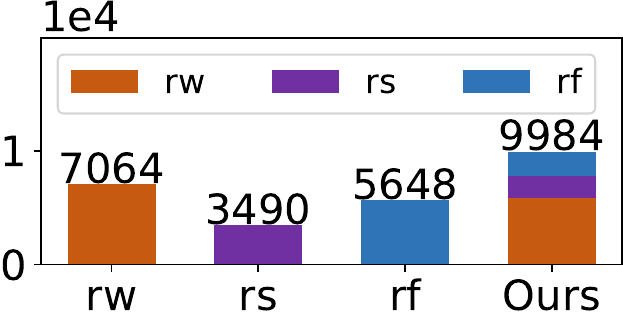}
        \caption{b18\_1}
    \end{subfigure}
    \hfill
    \begin{subfigure}[b]{0.19\textwidth}
        \centering
        \includegraphics[width=1\textwidth]{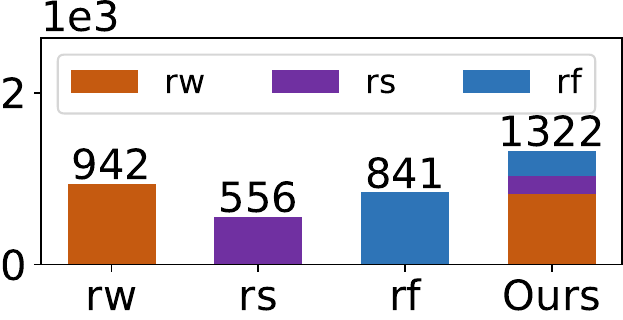}
        \caption{b21}
    \end{subfigure}
    \hfill
    \begin{subfigure}[b]{0.2\textwidth}
        \centering
        \includegraphics[width=1\textwidth]{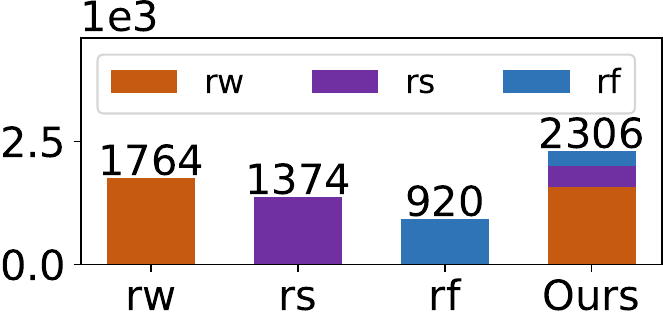}
        \caption{bfly}\label{fig:bfly_bd}
    \end{subfigure}
    \hfill
    \begin{subfigure}[b]{0.2\textwidth}
        \centering
        \includegraphics[width=1\textwidth]{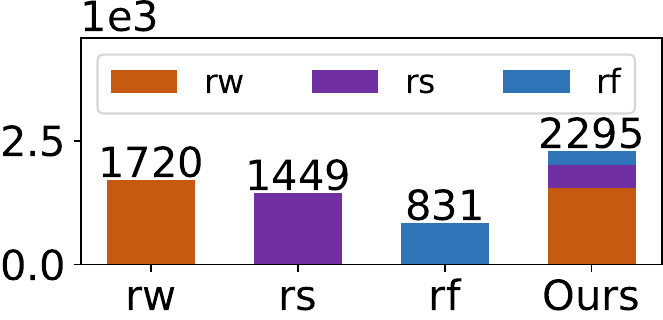}
        \caption{fir}
    \end{subfigure}
    \hfill
    \begin{subfigure}[b]{0.19\textwidth}
        \centering
        \includegraphics[width=1\textwidth]{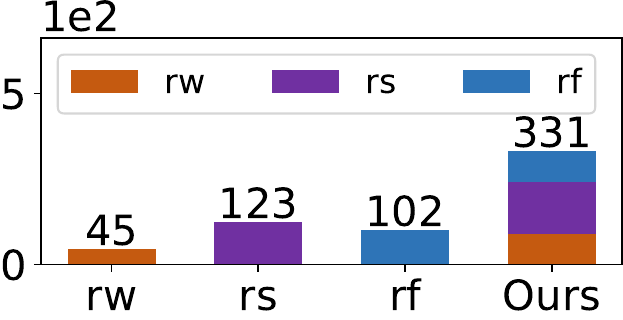}
        \caption{mem\_ctrl}
    \end{subfigure}
    \hfill
    \begin{subfigure}[b]{0.19\textwidth}
        \centering
        \includegraphics[width=1\textwidth]{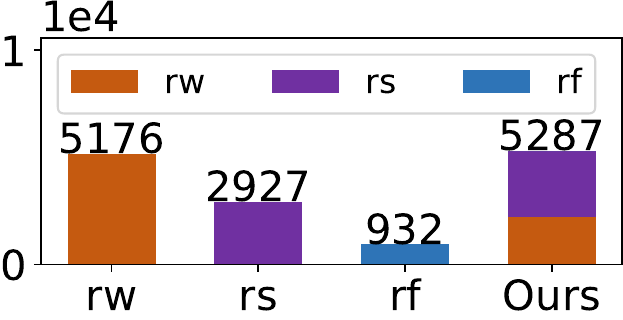}
        \caption{sqrt}
    \end{subfigure}
    \hfill
    \begin{subfigure}[b]{0.19\textwidth}
        \centering
        \includegraphics[width=1\textwidth]{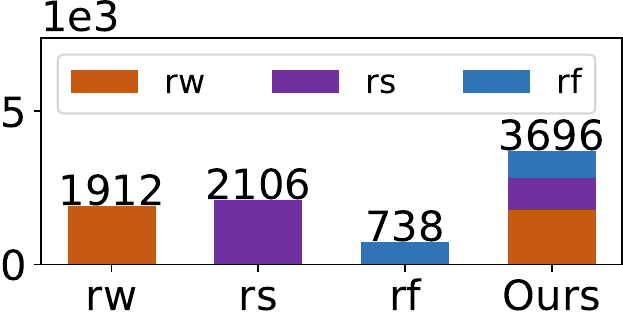}
        \caption{voter}
        \label{fig:ana_voter}
    \end{subfigure}
     \caption{{The optimization opportunities with different optimization operations. The $X-$axis denotes the optimization operations. The $Y-$axis denotes the number of valid iterations with the corresponding operation.}}
    % \vspace{-2mm}
     \label{fig:ana}
\end{figure*}

\begin{figure*}
    \centering
    %     \begin{subfigure}[b]{0.32\textwidth}
    % \centering
    %     \includegraphics[width=0.8\textwidth]{figs/tcad_venn_rw_rs_bfly.pdf}
    %     \caption{Venn diagram between \texttt{rw} and \texttt{rs}}
    %     \label{fig:venn_rw_rs}
    % \end{subfigure}
    % \hfill
    % \begin{subfigure}[b]{0.32\textwidth}
    % \centering
    %     \includegraphics[width=0.8\textwidth]{figs/tcad_venn_rw_rf_bfly.pdf}
    %     \caption{Venn diagram between \texttt{rw} and \texttt{rf}}
    %     \label{fig:venn_rw_rf}
    % \end{subfigure}
    % \hfill
    \begin{subfigure}[b]{0.24\textwidth}
    \centering
    \includegraphics[width=0.8\textwidth]{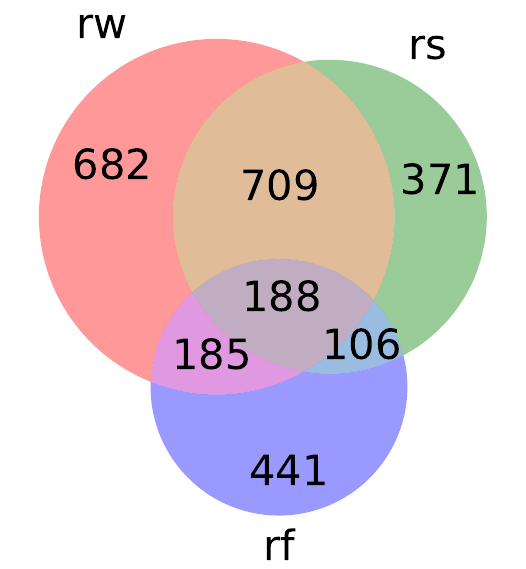}
        \caption{\texttt{rw} and \texttt{rs} and \texttt{rf}}
        \label{fig:venn_rw_rs_rf}
    \end{subfigure}
    \begin{subfigure}[b]{0.24\textwidth}
    \centering
        \includegraphics[width=0.8\textwidth]{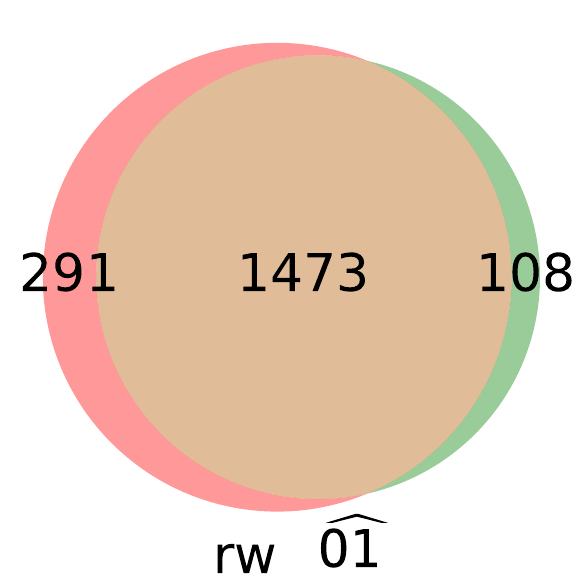}
        \caption{\texttt{rw} and the \texttt{rw} in $\widehat{\texttt{O1}}$}
        \label{fig:venn_o1_rw}
    \end{subfigure}
    \hfill
    \begin{subfigure}[b]{0.24\textwidth}
    \centering
        \includegraphics[width=0.8\textwidth]{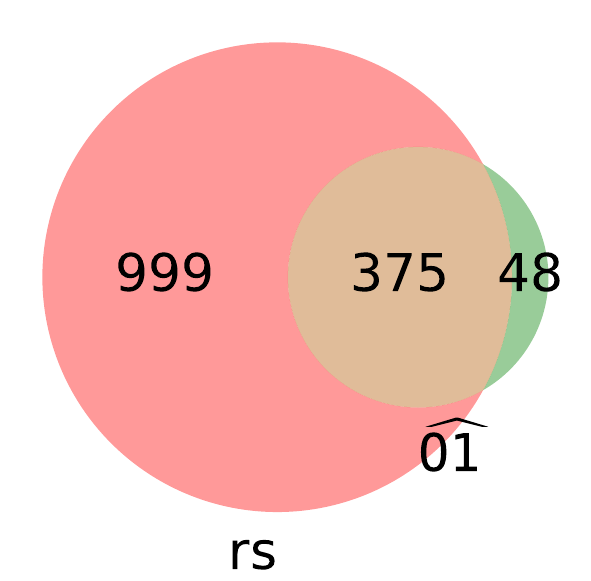}
        \caption{\texttt{rs} and the \texttt{rs} in $\widehat{\texttt{O1}}$}
        \label{fig:venn_o1_rs}
    \end{subfigure}
    \hfill
    \begin{subfigure}[b]{0.24\textwidth}
    \centering
\includegraphics[width=0.8\textwidth]{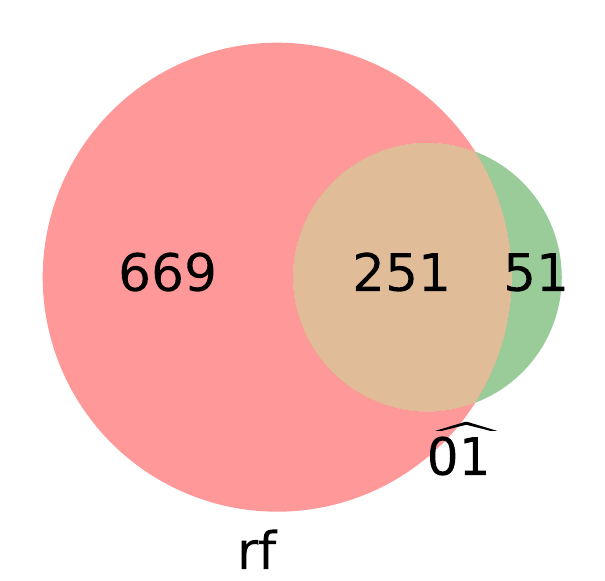}
        \caption{\texttt{rf} and the \texttt{rf} in $\widehat{\texttt{O1}}$}
        \label{fig:venn_o1_rf}
    \end{subfigure}
    
    \caption{{The Venn diagram for the design \textit{bfly}—a detailed analysis of Figure \ref{fig:bfly_bd}—illustrates the relationships as follows: (a) among standalone optimizations; and (b) to (d) between orchestration optimization $\widehat{\texttt{O1}}$ (as defined in Section \ref{sec: Orchrules}) and each of \texttt{rw}, \texttt{rs}, \texttt{rf}, respectively. In each diagram, we consider only the nodes that participate in valid iterations within the respective optimizations. Furthermore, for $\widehat{\texttt{O1}}$ in diagrams (b) to (d), the number of valid iterations is aligned with that of the corresponding optimization for comparison.}
    }
    \label{fig:venn}
\end{figure*}

{In existing logic optimization algorithms that follow a stand-alone optimization approach as shown in Figure \ref{fig:aig_alone}, certain nodes may miss optimization opportunities. For instance, node $g$, which is suitable for both \textit{refactoring} and \textit{resubstitution}, may be overlooked for optimizations in \textit{rewriting}.}
{To further enhance the logic optimization process in DAG-aware logic synthesis, we introduce "\textit{Orchestration}" for logic optimization in this work. This approach is in contrast to \textit{Stand-alone Logic Optimization} defined in Definition 1. We provide the details of \textit{Orchestrated Logic Optimization} in Definition 2.}

\noindent
{
\textbf{Definition 2: \textit{Orchestrated} Logic Optimization:} Orchestrated logic optimization involves multiple optimization operations being considered during a single traversal of the logic graph. {In each optimization iteration, multiple operations can be evaluated and applied based on the predefined orchestration criteria.}}

{In the orchestration optimization, multiple optimizations are made available for each node, thereby maximizing its optimization opportunities. Specifically, we orchestrate optimization operations including \textit{rewriting} (\texttt{rw}), \textit{resubstitution} (\texttt{rs}), and \textit{refactoring} (\texttt{rf}), in a single traversal of the AIG for the logic optimization. }The orchestration technique can be iteratively applied to the AIG multiple times, in combination with other optimization operations such as \textit{balance}, \textit{redundancy removal} to achieve iterative DAG optimization. Moreover, the optimized AIG resulting from our orchestration method can be verified for equivalence to the original AIG using Combinatorial Equivalence Checking (CEC). 

In this section, we first explore optimization opportunities in the single traversal of AIG for both stand-alone optimizations and orchestrated optimization. We then introduce two orchestration policies: \textit{Local-greedy orchestration}, which selects the operation yielding the highest local gain (i.e., the number of nodes saved by applying the optimization operation) at each node for AIG optimization, and \textit{Priority-ordered orchestration}, which prioritizes operations in a predefined order for AIG optimization at each node. %\yingjie{Finally, we provide an analysis of the runtime costs associated with the orchestration algorithms for logic optimization and compare them with existing stand-alone algorithms.}

\subsection{Optimization Opportunities Studies}
\label{sec:space}

First, we analyze the optimization opportunities in a single traversal of the AIG for various optimization methods. {We record the number of iterations where optimization leads to graph updates, termed as "valid iterations," in this analysis. The results for logic optimizations using \texttt{rw}, \texttt{rs}, \texttt{rf}, and the orchestration method are illustrated in Figure \ref{fig:ana}. The orange bar represents the number of valid iterations with \texttt{rw}, purple for \texttt{rs}, and blue for \texttt{rf}. The bar labeled "Ours" shows the number of valid iterations with the orchestration optimization, incorporating valid iterations from different optimizations (\texttt{rw}, \texttt{rs}, \texttt{rf}), indicated by the corresponding colors within the bar.} For instance, for the design \textit{voter}, stand-alone optimization methods (\texttt{rw}/\texttt{rs}/\texttt{rf}) result in $1917$/$2106$/$738$ valid iterations respectively (Figure \ref{fig:ana_voter}). In contrast, the orchestration method yields $3696$ valid iterations, presenting 93\%, 75\%, and 400\% more valid iterations than the \texttt{rw}, \texttt{rs}, and \texttt{rf} methods, respectively, in a single traversal of the AIG.

{For a better illustration, we present a Venn diagram using design \textit{bfly} in Figure \ref{fig:venn} as a detailed analysis of Figure \ref{fig:bfly_bd}. Here, the orchestration algorithm employed is the \textit{priority-ordered} algorithm with $\widehat{\texttt{O1}}$, which prioritizes \texttt{rw} most, then \texttt{rs} and \texttt{rf} least, and follows the definition in Section \ref{sec: Orchrules}. The diagram in Figure \ref{fig:venn_rw_rs_rf} demonstrates that while there are overlaps between different stand-alone optimizations, most root nodes found are distinct for each method. Additionally, the diagrams in Figure \ref{fig:venn_o1_rw} to \ref{fig:venn_o1_rf} for orchestration optimization and its corresponding stand-alone optimizations highlight unique root nodes in both approaches. It is noteworthy that the ratio of overlap to uniqueness varies with different orchestration algorithms and across designs.}

Our observations from this study highlight two key points: \textbf{(1)} Stand-alone optimization algorithms can miss significant optimization opportunities; \textbf{(2)} Orchestrating multiple optimizations in a single DAG traversal can introduce more optimization opportunities and more efficient logic optimization. %\yingjie{\textbf{(3)} For most designs, a large part of valid iterations with the orchestration optimization is contributed by \texttt{rw} (e.g., for designs \textit{bfly} and \textit{fir}).} 

{Given the context of orchestration, we can define the theoretical solution space and its optimal solution as follows: Consider a combinational And-Inverter Graph (AIG), denoted as \( G(V,E) \). It is postulated that within the entire solution space, which encompasses \( 3^{|V|} \) possibilities, there exists at least one orchestration decision ensuring that \( G(V,E) \) can be minimized to its smallest possible form utilizing a single traversal algorithm. Consequently, the theoretical upper limit for the complexity associated with pinpointing the optimal orchestration solution scales exponentially with the size of the graph, represented by \( |V| \). Nevertheless, within the scope of Boolean minimization, it has been empirically observed that the expansive solution space of \( 3^{|V|} \) may actually equate to a considerably reduced space of quality-of-results, specifically concerning the dimensions of the optimized AIGs. Note the solution space will increase if orchestration elaborates more than three synthesis techniques (i.e., increasing the base of the exponential complexity). This space of results tends to be notably constricted for smaller Boolean networks.}

Thus, to orchestrate multiple optimizations in a single AIG traversal, an effective orchestration policy (heuristic) is essential. In this work, we propose two policies: \textbf{(1) The \textit{Local-greedy orchestration}}, which selects the optimization operation resulting in the highest local gain (node reductions from the logic transformation of the operation) at the node for AIG optimization; and \textbf{(2) The \textit{Priority-ordered orchestration}}, which follows a pre-defined priority order for orchestrating multiple operations, i.e., applying optimizations according to the order. These policies are detailed in Algorithms \ref{alg:local} and \ref{alg:orch}, respectively.

\begin{figure}
\label{fig:aig_orch}
\centering
 \begin{subfigure}[b]{0.23\textwidth}
    \centering
        \includegraphics[width=1\textwidth]{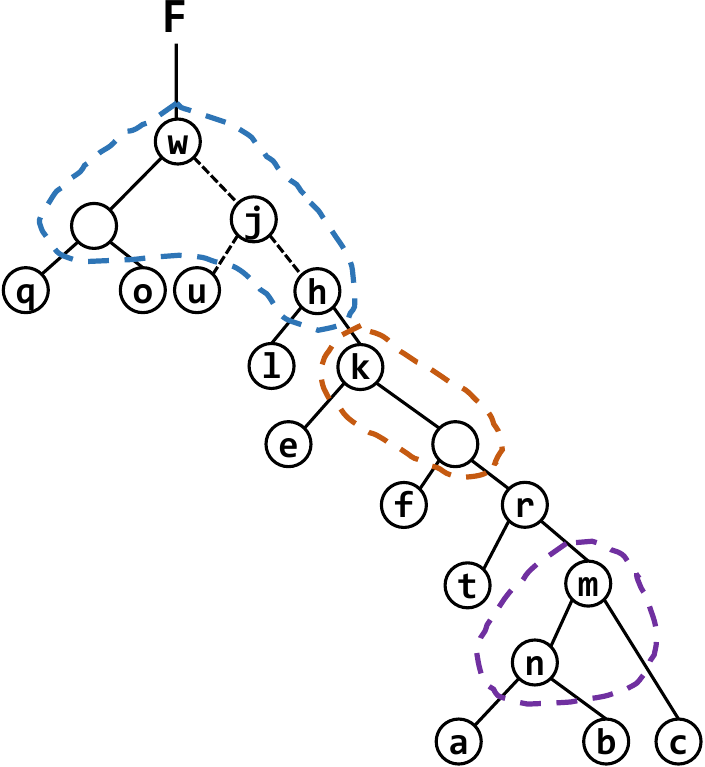}
        \caption{AIG with \texttt{Local-greedy} orchestration}
        \label{fig:aig_local}
    \end{subfigure}
    \hfill
    \begin{subfigure}[b]{0.23\textwidth}
    \centering
        \includegraphics[width=1\textwidth]{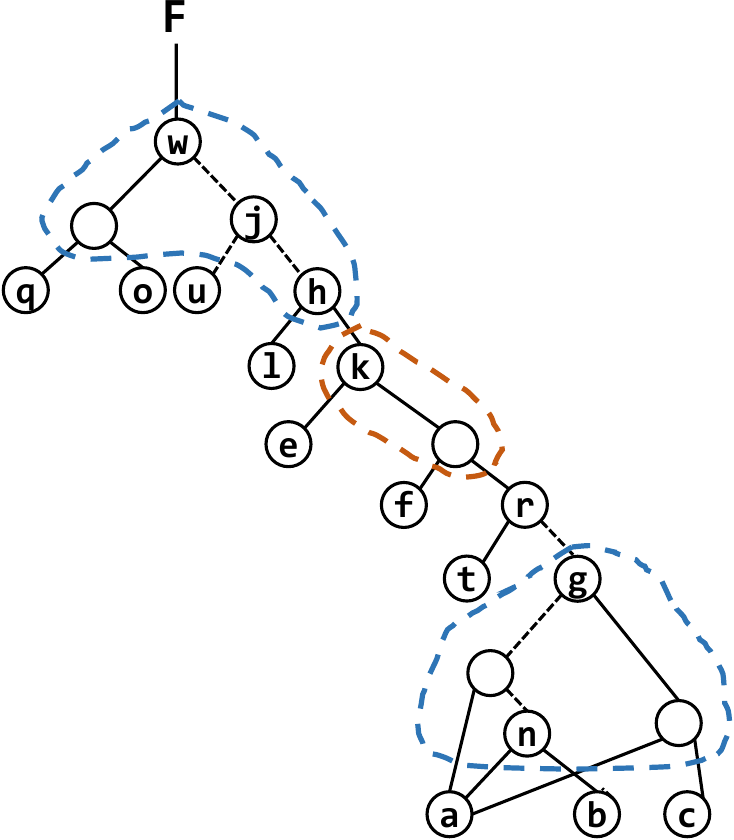}
        \caption{AIG with \texttt{Priority-ordered} orchestration}
        \label{fig:aig_pri}
    \end{subfigure}
    \caption{{The optimized graph produced by orchestration optimization operations: (a) optimized AIG with \textit{Local-greedy orchestration}, graph size is 19; (b) optimized AIG with \textit{Priority-ordered orchestration}, graph size is 21.}}
\end{figure}

\subsection{{Algorithm 1: Local-greedy Orchestration}}
The \textit{Local-greedy orchestration} algorithm takes a graph $G(V, E)$ as input, where $V$ represents the set of nodes in the AIG, and $E$ denotes the edges between nodes. Following the topological order, we initially check the transformability of each node with respect to all orchestrated optimization operations, namely \texttt{rw}, \texttt{rs}, and \texttt{rf}. This process yields the corresponding local optimization gains, $G_{rw}$, $G_{rs}$, and $G_{rf}$ (line 2). When none of the operations are applicable to a node, the local gain $G$ is set to $-1$.
{Once the local gains $G_{rw}$, $G_{rs}$, and $G_{rf}$ at the node are determined, the algorithm identifies the optimization operation with the highest non-negative local gain (lines 3, 6, and 9). The operation with the highest gain is then applied, and the graph is updated accordingly (lines 4, 7, and 10). If no operation is applicable (all gains are negative), the node is bypassed for optimization (line 12), and the algorithm proceeds to the next node in the iteration (line 13).}

{Compared to \textit{stand-alone optimizations}, the \textit{Local-greedy} orchestration algorithm incurs additional runtime overhead due to the necessity of pre-computing transformability checks and local gains for all available optimization operations (line 2).}%However, this overhead is offset by the reduced number of required iterations resulting from the effective orchestration of optimizations. This trade-off between overhead and efficiency is further discussed in Section \ref{sec:runtime_result}.}

\begin{algorithm}
\caption{Local-greedy Orchestration}   
\footnotesize
\label{alg:local}
 
   \SetKwInOut{KwIn}{Input}
    \SetKwInOut{KwOut}{Output}
    \KwIn{~$G(V,E) \leftarrow$ Boolean Networks/Circuits in AIG} 
    \KwOut{~Post-optimized AIG $G(V,E)$}
 \For{$v \in V$ in topological order}{
 %\tcp{New order}
       check transformability of $v$ w.r.t orchestrated operations: \texttt{rw}, \texttt{rs}, \texttt{rf}, and get the corresponding optimization gain: \texttt{G$_{rw}^{v}$}, \texttt{G$_{rs}^{v}$}, \texttt{G$_{rf}^{v}$}.
        \tcp*[f]{if operation is not applicable, $G$ is $-1$; otherwise, $G$ is a non-negative number.}

  \uIf{\texttt{G$_{rw}^{v}$} $\geq$ 0 and \texttt{G$_{rw}^{v}$} $\geq$ \texttt{G$_{rs}^{v}$} and \texttt{G$_{rw}^{v}$} $\geq$ \texttt{G$_{rf}^{v}$} }{
  Apply ${rw}$ to $v$ and update $G(V,E)$ \\ %$ \leftarrow$ \texttt{Dec\_GraphUpdateNetwork} \\
        continue
     \tcp*[f]{\texttt{rw} with the highest gain}
         }
  \uElseIf{\texttt{G$_{rs}^{v}$} $\geq$ 0 and \texttt{G$_{rs}^{v}$} $\geq$ \texttt{G$_{rw}^{v}$} and \texttt{G$_{rs}^{v}$} $\geq$ \texttt{G$_{rf}^{v}$} }{
  Apply ${rs}$ to $v$ and update $G(V,E)$ \\ %$ \leftarrow$ \texttt{Dec\_GraphUpdateNetwork} \\   
        continue
     \tcp*[f]{\texttt{rs} with the highest gain}
  }
  \uElseIf{\texttt{G$_{rf}^{v}$} $\geq$ 0 and \texttt{G$_{rf}^{v}$} $\geq$ \texttt{G$_{rw}^{v}$} and \texttt{G$_{rf}^{v}$} $\geq$ \texttt{G$_{rs}^{v}$} }{
  Apply ${rf}$ to $v$ and update $G(V,E)$ \\ %$ \leftarrow$ \texttt{Dec\_GraphUpdateNetwork} \\
        continue
     \tcp*[f]{\texttt{rf} with the highest gain}
  }
  \Else{
          %Update $G(V,E)$ \\ %$\leftarrow$ \texttt{Dec\_GraphUpdateNetwork} \\
    continue 
  }
  }

\end{algorithm}

\noindent
\textbf{Example:} An illustrative example is shown in Figure \ref{fig:aig_local}, based on the original AIG from Figure \ref{fig:aig_ori}. Following the topological order of the AIG, the Primary Inputs (PIs) are bypassed for optimization. Nodes $n$, $m$, $d$, and $p$ are also skipped as none of the optimizations are applicable to them. {The algorithm then evaluates node $g$, checking its transformability with \texttt{rw}, \texttt{rs}, and \texttt{rf}, and determining the local gains as $G_{rw} = -1$, $G_{rs} = 3$, and $G_{rf} = 1$. The \textit{Local-greedy orchestration} algorithm selects the operation with the highest local gain for optimization, in this case, \texttt{rs}. By iteratively traversing the entire logic graph, the \textit{Local-greedy orchestration} algorithm optimizes the AIG with a node reduction of $6$, as depicted in Figure \ref{fig:aig_local}.}

\subsection{{Algorithm 2: Priority-ordered Orchestration}} \label{sec: Orchrules}
In this algorithm, the selection of the optimization operation to be applied at each node depends on a pre-defined priority order with respect to the available optimizations. For the three operations \texttt{rw}, \texttt{rs}, and \texttt{rf}, there are six possible permutations of the priority order, namely: 
$\widehat{\texttt{O1}}$ $\mapsto$ \texttt{rw}$>$\texttt{rs}$>$\texttt{rf}, $\widehat{\texttt{O2}}$ $\mapsto$ \texttt{rw}$>$\texttt{rf}$>$\texttt{rs}, $\widehat{\texttt{O3}}$ $\mapsto$ \texttt{rs}$>$\texttt{rw}$>$\texttt{rf}, $\widehat{\texttt{O4}}$ $\mapsto$ \texttt{rs}$>$\texttt{rf}$>$\texttt{rw}, $\widehat{\texttt{O5}}$ $\mapsto$ \texttt{rf}$>$\texttt{rs}$>$\texttt{rw}, and $\widehat{\texttt{O6}}$ $\mapsto$ \texttt{rf}$>$\texttt{rw}$>$\texttt{rs}. {For instance, the priority order $\widehat{\texttt{O1}}$ implies that \texttt{rw} has the highest priority during optimization, meaning it is checked first for transformability; \texttt{rs} is evaluated next if \texttt{rw} is not applicable to the node; and \texttt{rf} is considered last when the higher priority operations are not applicable.}

%The proposed orchestration implementation is described in Algorithm \ref{alg:orch}. While traversing the AIG, the nodes are visited in topological order (line 1). Given the input AIG $G(V, E)$ and the priority orchestration policy $P_{>}$, where the first position has the highest priority, at the unseen node $v$, the algorithm checks the transformability of the operation at the node following the policy $P_{>}$. When the first priority operation is available, this operation will be selected and applied to update the graph, and the algorithm will move to the next iteration while ignoring transformability checks with other operations (lines 2 -- 4). If the transformability check for the first operation fails at the node, the transformability w.r.t the operation at the second position will be checked for the node. When the second operation is applicable, it will apply the second operation and update the graph accordingly and move to the next iteration (lines 5 -- 7). The algorithm will check the least priority operation when the first two operations are both inapplicable (lines 8 -- 10). When all the operations are checked, and none of them are applicable, the node will be skipped for the optimization, and the algorithm moves to the next iteration without graph updates (lines 11 -- 13). Note that the the leave nodes in the optimized cuts of $v$ are marked as the seen nodes, which are excluded from the successive iterations (line 14). The algorithm completes the optimization for each node in the AIG graph iteratively until no unseen node exists in the updated graph.  

{Algorithm \ref{alg:orch} outlines the implementation of the priority-ordered orchestration for logic graphs. This algorithm takes an AIG $G(V, E)$ and a priority orchestration policy $P_{>}$ as inputs. The policy $P_{>}$ is defined as a precedence-ordered set of operations, wherein the operation positioned first has the highest priority. 
%In each iteration at node $v$, the algorithm assesses the feasibility of performing an operation following the policy $P_{>}$. Upon discovering a viable operation following the priority order, the operation is applied to the current node and update the graph, and the algorithm will exit the current iteration and move to the next iteration without evaluations for other available operations with lower priority. 
{As delineated in Algorithm \ref{alg:orch}, for each node, the algorithm initially examines the transformability of the highest-priority operation, $P_{>}[0]$ (line 2). If $P_{>}[0]$ is applicable, it is applied, and the graph is updated accordingly (line 3). Following this, the algorithm proceeds to the next node without evaluating other lower-priority operations (line 4).
If $P_{>}[0]$ is not applicable, the algorithm assesses the next highest-priority operation, $P_{>}[1]$ (lines 5 -- 7). This process is repeated, methodically evaluating operations in descending order of priority (lines 8 -- 10). In cases where none of the operations in the policy $P_{>}$ are applicable, the node is bypassed in the current iteration, resulting in no modifications to the graph (lines 11 -- 12).}
}

{The selection of the most effective priority order depends heavily on the specific design domain. The initial transformation chosen can significantly impact the optimization process. Operations with higher priority tend to play a more critical role. Furthermore, incorporating domain knowledge into the optimization process can improve performance. Machine learning techniques can be helpful in this regard and exploring their potential leads to more further work.}

\begin{algorithm}
\caption{Priority-ordered orchestration}   
%\begin{algorithmic} %\begin{algorithmic}[1]
%\vspace{-2mm}
\footnotesize
\label{alg:orch}
 
   \SetKwInOut{KwIn}{Input}
     \SetKwInOut{KwIn}{Input}
    \SetKwInOut{KwOut}{Output}
    \KwIn{~$G(V,E) \leftarrow$ Boolean Networks/Circuits in AIG} 
    \KwIn{~Orchestration rule: $\mathcal{P}_>$(\texttt{rw, rf, rs}) }
    \tcp{$\mathcal{P}_>$ is a list as the permutation of the available Boolean transformations.}
    \KwOut{~Post-optimized AIG $G(V,E)$}
 \For{$v \in V$ in topological order}{
 %\tcp{New order}
  \uIf{v \text{is transformable w.r.t} $\mathcal{P}_>$[0]}{
        Apply ${\mathcal{P}_>[0]}$ to $v$ and update $G(V,E) $\\%$ \leftarrow$ \texttt{Dec\_GraphUpdateNetwork} \\
        continue
     \tcp*[f]{check first priority}
         }
  \uElseIf{v \text{is transformable w.r.t} $\mathcal{P}_>$[1]}{
        Apply ${\mathcal{P}_>[1]}$ to $v$ and update $G(V,E)$\\%$ \leftarrow$ \texttt{Dec\_GraphUpdateNetwork} \\
        continue
     \tcp*[f]{check second priority}
  }
  \uElseIf{v \text{is transformable w.r.t} $\mathcal{P}_>$[2]}{
        Apply ${\mathcal{P}_>[2]}$ to $v$ and update $G(V,E) $\\%$ \leftarrow$ \texttt{Dec\_GraphUpdateNetwork} \\
        continue
     \tcp*[f]{check third priority}
  }
  \Else{
    %Update $G(V,E) \leftarrow$ \texttt{Dec\_GraphUpdateNetwork} \\
    continue 
  }
  %Update $G(V,E) \leftarrow$ \texttt{Dec\_GraphUpdateNetwork}, and exclude $v$ from $V$
  }
%  \end{algorithmic}
\end{algorithm}

\iffalse
\begin{algorithm}
\caption{Pseudo-code for synthesis orchestration}\label{alg:orch}
\textbf{procedure} DAG-Aware Synthesis Orchestration()\\
$G(V, E)$ $\leftarrow$ Boolean Circuits/Network to AIG \\
\For{v $\in$ V } %\COMMENT{this is a comment}
{    
   \texttt{valid\_ops}=\{ \}; $\leftarrow$ applicable operations for the node \\
   check transformability for all \textit{rw}, \textit{rs}, \textit{rf} statically; \\
   collect transformable nodes in \texttt{valid\_ops}; \\
   pick the operation based on the priority sorting \texttt{orch}; \\
   %select_operation(v) = orch(valid\_ops);\\
    \If{transformable(v)}
    {
        apply transformation to $v$;
    }
    update G(V, E)
}
\end{algorithm}
\fi 

\noindent
\textbf{Example:} An illustrative example is shown in Figure \ref{fig:aig_pri}, using the original AIG from Figure \ref{fig:aig_ori}. The priority sorting is set as $P_{>}(\texttt{rw}, \texttt{rf}, \texttt{rs})$, corresponding to $\widehat{O2}$. Following the topological order, the PIs and nodes $n$, $m$, $d$, and $p$ are bypassed as none of the optimizations are applicable. For node $g$, the algorithm first checks \texttt{rw} as per the priority order but finds it inapplicable, proceeding then to \texttt{rf}. Since \texttt{rf} is applicable, it is applied to update the AIG (indicated by the blue box), with no need to check the transformability of \texttt{rs}. {The iterative traversal of the entire graph leads to the AIG, optimized via the \textit{Priority-ordered orchestration} algorithm, achieving $4$ node reductions compared to the original graph.}

{These two orchestration algorithms apply stand-alone optimizations within a single AIG traversal, each leveraging distinct strategies. The \textit{Local-greedy orchestration} algorithm selects the most effective operation for logic minimization based on the current local node structure. In contrast, the \textit{Priority-ordered orchestration} algorithm utilizes a variety of pre-defined priority orders, potentially enhancing overall performance. A key distinction lies in their operational approach: the \textit{Local-greedy orchestration} algorithm examines the transformability with respect to all operations at each node, whereas the \textit{Priority-ordered} algorithm progresses to the next node once an applicable operation is found in the given order, effectively minimizing redundant transformability checks. Consequently, in terms of runtime efficiency, the \textit{Local-greedy orchestration} may be less efficient compared to the \textit{Priority-ordered orchestration}. Detailed empirical studies and discussions of these findings are presented in Section \ref{sec:experiment}.}

%\cy{I am thinking about adding a short self-discussion here to quickly discuss the differences between algorithms 1 and 2, which connects to our experimental setups in next section. For example, local greedy with have no problem of exploring any algorithm space (does not have any) but orchestration has. If has options like priority orch, then the flow problem will exsit. anything else?}

%As a result, compared with the stand-alone optimizations produced by ABC in Figure \ref{fig:aig_rw} -- Figure \ref{fig:aig_rs}, the \texttt{orchestration}-optimized graph shown in Figure \ref{fig:aig_orch} shows better logic minimization optimization performance in fewer optimization iterations as shown in Table \ref{tbl:reduce}. With skipped nodes in the stand-alone operations optimized with other orchestrated operations, more nodes within the operation cuts are marked "seen" nodes in each iteration, thus highly reducing the iteration numbers and improving the minimization performance.

\begin{table*}[t]
\centering
\caption{Detailed results of selected large size designs. Comparison of single-traversal \textit{orchestration} with stand-alone optimizations from ABC.}
\resizebox{\textwidth}{!}{%
\begin{tabular}{|ccccccccccccc|}
\hline
\multicolumn{2}{|c|}{\multirow{3}{*}{Design}} & \multicolumn{11}{c|}{AIG} \\ \cline{3-13} 
\multicolumn{2}{|c|}{} & \multicolumn{1}{c|}{Baseline} & \multicolumn{1}{c|}{\texttt{rw}} & \multicolumn{1}{c|}{\texttt{rs}} & \multicolumn{1}{c|}{\texttt{rf}} & \multicolumn{1}{c|}{$\widehat{\texttt{O1}}$} & \multicolumn{1}{c|}{$\widehat{\texttt{O2}}$} & \multicolumn{1}{c|}{$\widehat{\texttt{O3}}$} & \multicolumn{1}{c|}{$\widehat{\texttt{O4}}$} & \multicolumn{1}{c|}{$\widehat{\texttt{O5}}$} & \multicolumn{1}{c|}{$\widehat{\texttt{O6}}$} & ${\texttt{LocalGreedy}}$\\ \cline{3-13} 
\multicolumn{2}{|c|}{} & \multicolumn{1}{c|}{\#Node} & \multicolumn{1}{c|}{\#Node ($\Delta$\%)} & \multicolumn{1}{c|}{\#Node ($\Delta$\%)} & \multicolumn{1}{c|}{\#Node ($\Delta$\%)} & \multicolumn{1}{c|}{\#Node ($\Delta$\%)} & \multicolumn{1}{c|}{\#Node ($\Delta$\%)} & \multicolumn{1}{c|}{\#Node ($\Delta$\%)} & \multicolumn{1}{c|}{\#Node ($\Delta$\%)} & \multicolumn{1}{c|}{\#Node ($\Delta$\%)} & \multicolumn{1}{c|}{\#Node ($\Delta$\%)} & \#Node ($\Delta$\%) \\ \hline
\multicolumn{1}{|c|}{\multirow{7}{*}{ISCAS}} & \multicolumn{1}{c|}{s38584} & \multicolumn{1}{c|}{12400} & \multicolumn{1}{c|}{10697 (13.7\%)} & \multicolumn{1}{c|}{11505 (7.2\%)} & \multicolumn{1}{c|}{10932 (11.8\%)} & \multicolumn{1}{c|}{10366 (16.4\%)} & \multicolumn{1}{c|}{10379 (16.3\%)} & \multicolumn{1}{c|}{\textbf{10336 (16.7\%)}} & \multicolumn{1}{c|}{10655 (14.1\%)} & \multicolumn{1}{c|}{10932 (11.8\%)} & \multicolumn{1}{c|}{10932 (11.8\%)} & 10449 (15.7\%) \\ \cline{2-13} 
\multicolumn{1}{|c|}{} & \multicolumn{1}{c|}{s35932} & \multicolumn{1}{c|}{11948} & \multicolumn{1}{c|}{9110 (23.8\%)} & \multicolumn{1}{c|}{11916 (0.3\%)} & \multicolumn{1}{c|}{9836 (17.7\%)} & \multicolumn{1}{c|}{\textbf{8561 (28.4\%)}} & \multicolumn{1}{c|}{\textbf{8561 (28.4\%)}} & \multicolumn{1}{c|}{\textbf{8561 (28.4\%)}} & \multicolumn{1}{c|}{9836 (17.7\%)} & \multicolumn{1}{c|}{9836 (17.7\%)} & \multicolumn{1}{c|}{9836 (17.7\%)} & \textbf{8561 (28.4\%)} \\ \cline{2-13} 
\multicolumn{1}{|c|}{} & \multicolumn{1}{c|}{b17\_1} & \multicolumn{1}{c|}{27647} & \multicolumn{1}{c|}{24178 (12.6\%)} & \multicolumn{1}{c|}{26305 (4.9\%)} & \multicolumn{1}{c|}{24533 (11.3\%)} & \multicolumn{1}{c|}{22951 (17.0\%)} & \multicolumn{1}{c|}{23125 (16.4\%)} & \multicolumn{1}{c|}{\textbf{22935 (17.0\%)}} & \multicolumn{1}{c|}{23393 (15.4\%)} & \multicolumn{1}{c|}{24532 (11.3\%)} & \multicolumn{1}{c|}{24532 (11.3\%)} & {23008 (16.7\%)} \\ \cline{2-13} 
\multicolumn{1}{|c|}{} & \multicolumn{1}{c|}{b18\_1} & \multicolumn{1}{c|}{79054} & \multicolumn{1}{c|}{66807 (15.5\%)} & \multicolumn{1}{c|}{73076 (7.6\%)} & \multicolumn{1}{c|}{69606 (12.0\%)} & \multicolumn{1}{c|}{63431 (19.8\%)} & \multicolumn{1}{c|}{64135 (18.9\%)} & \multicolumn{1}{c|}{\textbf{63167 (20.1\%)}} & \multicolumn{1}{c|}{65956 (16.6\%)} & \multicolumn{1}{c|}{69586 (12.0\%)} & \multicolumn{1}{c|}{69587 (12.0\%)} & 63726 (19.4\%) \\ \cline{2-13} 

\multicolumn{1}{|c|}{} & \multicolumn{1}{c|}{b20} & \multicolumn{1}{c|}{12219} & \multicolumn{1}{c|}{10659 (12.8\%)} & \multicolumn{1}{c|}{11197 (8.4\%)} & \multicolumn{1}{c|}{10593 (13.3\%)} & \multicolumn{1}{c|}{10017 (18.02\%)} & \multicolumn{1}{c|}{10110 (17.3\%)} & \multicolumn{1}{c|}{\textbf{10011 (18.1\%)}} & \multicolumn{1}{c|}{10228 (16.3\%)} & \multicolumn{1}{c|}{10590 (13.3\%)} & \multicolumn{1}{c|}{10590 (13.3\%)} & 10129 (17.1\%) \\ \cline{2-13} 

\multicolumn{1}{|c|}{} & \multicolumn{1}{c|}{b21} & \multicolumn{1}{c|}{12782} & \multicolumn{1}{c|}{10863 (15.1\%)} & \multicolumn{1}{c|}{11449 (10.4\%)} & \multicolumn{1}{c|}{10961 (14.3\%)} & \multicolumn{1}{c|}{10146 (20.6\%)} & \multicolumn{1}{c|}{10237 (19.9\%)} & \multicolumn{1}{c|}{\textbf{10133 (20.7\%)}} & \multicolumn{1}{c|}{10458 (18.2\%)} & \multicolumn{1}{c|}{10958 (14.3\%)} & \multicolumn{1}{c|}{10958 (14.3\%)} & 10261 (19.7\%) \\ \cline{2-13} 

\multicolumn{1}{|c|}{} & \multicolumn{1}{c|}{b22} & \multicolumn{1}{c|}{18488} & \multicolumn{1}{c|}{15983 (13.6\%)} & \multicolumn{1}{c|}{16891 (8.6\%)} & \multicolumn{1}{c|}{15983 (13.6\%)} & \multicolumn{1}{c|}{14977 (19.0\%)} & \multicolumn{1}{c|}{15115 (18.2\%)} & \multicolumn{1}{c|}{\textbf{14953 (19.1\%)}} & \multicolumn{1}{c|}{15275 (17.4\%)} & \multicolumn{1}{c|}{15965 (13.6\%)} & \multicolumn{1}{c|}{15965 (13.6\%)} & 15127 (18.2\%) \\ \hline

\multicolumn{1}{|c|}{\multirow{4}{*}{VTR}} & \multicolumn{1}{c|}{bfly} & \multicolumn{1}{c|}{28910} & \multicolumn{1}{c|}{26827 (7.2\%)} & \multicolumn{1}{c|}{27060 (6.4\%)} & \multicolumn{1}{c|}{27487 (4.9\%)} & \multicolumn{1}{c|}{\textbf{25996 (10.1\%)}} & \multicolumn{1}{c|}{26183 (9.4\%)} & \multicolumn{1}{c|}{26027 (10.0\%)} & \multicolumn{1}{c|}{26353 (8.8\%)} & \multicolumn{1}{c|}{27487 (4.9\%)} & \multicolumn{1}{c|}{27487 (4.9\%)} & 26181 (9.4\%) \\ \cline{2-13} 

\multicolumn{1}{|c|}{} & \multicolumn{1}{c|}{dscg} & \multicolumn{1}{c|}{28252} & \multicolumn{1}{c|}{26132 (7.5\%)} & \multicolumn{1}{c|}{26352 (6.73\%)} & \multicolumn{1}{c|}{26972 (4.5\%)} & \multicolumn{1}{c|}{\textbf{25339 (10.3\%)}} & \multicolumn{1}{c|}{25552 (9.5\%)} & \multicolumn{1}{c|}{25345 (10.3\%)} & \multicolumn{1}{c|}{25768 (8.8\%)} & \multicolumn{1}{c|}{26970 (4.5\%)} & \multicolumn{1}{c|}{26970 (4.5\%)} & {25496 (9.7\%)} \\ \cline{2-13} 

\multicolumn{1}{|c|}{} & \multicolumn{1}{c|}{fir} & \multicolumn{1}{c|}{27704} & \multicolumn{1}{c|}{25641 (7.5\%)} & \multicolumn{1}{c|}{25768 (7.0\%)} & \multicolumn{1}{c|}{26437 (4.6\%)} & \multicolumn{1}{c|}{\textbf{24778 (10.6\%)}} & \multicolumn{1}{c|}{25061 (9.5\%)} & \multicolumn{1}{c|}{24831 (10.4\%)} & \multicolumn{1}{c|}{25189 (9.1\%)} & \multicolumn{1}{c|}{26437 (4.6\%)} & \multicolumn{1}{c|}{26437 (4.6\%)} & 24987 (9.8\%) \\ \cline{2-13}

\multicolumn{1}{|c|}{} & \multicolumn{1}{c|}{syn2} & \multicolumn{1}{c|}{30003} & \multicolumn{1}{c|}{27787 (7.4\%)} & \multicolumn{1}{c|}{28031 (6.6\%)} & \multicolumn{1}{c|}{28617 (4.6\%)} & \multicolumn{1}{c|}{\textbf{27013 (10.0\%)}} & \multicolumn{1}{c|}{27266 (9.1\%)} & \multicolumn{1}{c|}{27048 (9.9\%)} & \multicolumn{1}{c|}{27444 (8.5\%)} & \multicolumn{1}{c|}{28617 (4.6\%)} & \multicolumn{1}{c|}{28617 (4.6\%)} & {27198 (9.3\%)} \\ \hline

\multicolumn{1}{|c|}{\multirow{5}{*}{EPFL}} & \multicolumn{1}{c|}{div} & \multicolumn{1}{c|}{57247} & \multicolumn{1}{c|}{41153 (28.1\%)} & \multicolumn{1}{c|}{52621 (8.1\%)} & \multicolumn{1}{c|}{56745 (0.9\%)} & \multicolumn{1}{c|}{\textbf{41123 (28.2\%)}} & \multicolumn{1}{c|}{41143 (28.1\%)} & \multicolumn{1}{c|}{41124 (28.2\%)} & \multicolumn{1}{c|}{52098 (9.0\%)} & \multicolumn{1}{c|}{56738 (0.9\%)} & \multicolumn{1}{c|}{56738 (0.9\%)} & 41147 (28.1\%) \\ \cline{2-13} 

\multicolumn{1}{|c|}{} & \multicolumn{1}{c|}{hyp} & \multicolumn{1}{c|}{214335} & \multicolumn{1}{c|}{214274 (0.0\%)} & \multicolumn{1}{c|}{209164 (2.4\%)} & \multicolumn{1}{c|}{212341 (1.0\%)} & \multicolumn{1}{c|}{207335 (3.3\%)} & \multicolumn{1}{c|}{212327 (0.9\%)} & \multicolumn{1}{c|}{\textbf{207315 (3.3\%)}} & \multicolumn{1}{c|}{\textbf{207315 (3.3\%)}} & \multicolumn{1}{c|}{212338 (1.0\%)} & \multicolumn{1}{c|}{212338 (1.0\%)} & 207319 (3.3\%) \\ \cline{2-13} 

\multicolumn{1}{|c|}{} & \multicolumn{1}{c|}{mem\_ctrl} & \multicolumn{1}{c|}{46836} & \multicolumn{1}{c|}{46732 (0.2\%)} & \multicolumn{1}{c|}{46554 (0.6\%)} & \multicolumn{1}{c|}{46574 (0.6\%)} & \multicolumn{1}{c|}{46301 (1.1\%)} & \multicolumn{1}{c|}{46484 (0.8\%)} & \multicolumn{1}{c|}{\textbf{46085 (1.6\%)}} & \multicolumn{1}{c|}{46204 (1.4\%)} & \multicolumn{1}{c|}{46569 (0.6\%)} & \multicolumn{1}{c|}{46569 (0.6\%)} & {46201 (1.3\%)} \\ \cline{2-13}

\multicolumn{1}{|c|}{} & \multicolumn{1}{c|}{sqrt} & \multicolumn{1}{c|}{24618} & \multicolumn{1}{c|}{19441 (21.0\%)} & \multicolumn{1}{c|}{21690 (11.9\%)} & \multicolumn{1}{c|}{23685 (3.8\%)} & \multicolumn{1}{c|}{\textbf{19221 (21.9\%)}} & \multicolumn{1}{c|}{19441 (21.0\%)} & \multicolumn{1}{c|}{\textbf{19221 (21.9\%)}} & \multicolumn{1}{c|}{21582 (12.3\%)} & \multicolumn{1}{c|}{23685 (3.8\%)} & \multicolumn{1}{c|}{23685 (3.8\%)} & 19221 (21.9\%) \\ \cline{2-13} 

\multicolumn{1}{|c|}{} & \multicolumn{1}{c|}{voter} & \multicolumn{1}{c|}{13758} & \multicolumn{1}{c|}{11408 (17.0\%)} & \multicolumn{1}{c|}{10997 (20.1\%)} & \multicolumn{1}{c|}{12681 (7.8\%)} & \multicolumn{1}{c|}{9461 (31.2\%)} & \multicolumn{1}{c|}{10982 (20.2\%)} & \multicolumn{1}{c|}{\textbf{9399 (31.7\%)}} & \multicolumn{1}{c|}{9492 (31.0\%)} & \multicolumn{1}{c|}{12679 (7.8\%)} & \multicolumn{1}{c|}{12679 (7.8\%)} & 9606 (30.2\%) \\ \hline

\multicolumn{3}{|c|}{\textbf{Avg. Node Reduction\%}} & \multicolumn{1}{c|}{12.7\%} & \multicolumn{1}{c|}{7.3\%} & \multicolumn{1}{c|}{7.9\%} & \multicolumn{1}{c|}{16.6\%} & \multicolumn{1}{c|}{15.3\%} & \multicolumn{1}{c|}{\textbf{16.7\%}} & \multicolumn{1}{c|}{13.0\%} & \multicolumn{1}{c|}{7.9\%} & \multicolumn{1}{c|}{7.9\%} & 16.1\% \\ \hline

\multicolumn{3}{|c|}{{\textbf{Avg. Runtime (s)}}} & \multicolumn{1}{c|}{{0.366}} & \multicolumn{1}{c|}{{0.177}} & \multicolumn{1}{c|}{{0.155}} & \multicolumn{1}{c|}{{0.459}} & \multicolumn{1}{c|}{{0.373}} & \multicolumn{1}{c|}{{0.454}} & \multicolumn{1}{c|}{{0.226}} & \multicolumn{1}{c|}{{0.123}} & \multicolumn{1}{c|}{{0.137}} & {0.478} \\ \hline
\end{tabular}
}
% \caption{Detailed results of selected large size designs. Comparison of single-traversal \texttt{orchestration} with stand-alone optimizations from ABC.}
% \vspace{-3mm}
\label{table:1O}
\end{table*}

\section{Experiments}
\label{sec:experiment}

%We evaluate the performance of the proposed \texttt{orchestration} synthesis in logic minimization and area reduction after post-synthesis technology-mapping, in single traversal optimization, iterative optimization, and synthesis flow settings. 
Our experimental results include comparisons with stand-alone optimizations in ABC, covering: \textbf{(1)} performance and runtime evaluation with single traversal optimization;
%\yingjie{and performance evaluation with single traversal of all available optimizations, where we apply single traversal of graph with each optimization (\texttt{rw}, \texttt{rs}, \texttt{rf}) for sequential optimization compared with the sequential optimization with three times of sequential orchestration optimization}; 
 \textbf{(2)} performance evaluation of optimization methods in iterative synthesis; \textbf{(3)} end-to-end performance evaluation in existing ABC flows \texttt{resyn} by OpenROAD \cite{ajayi2019openroad}, where the orchestration method is integrated into ABC in Yosys \cite{wolf2016yosys} for OpenROAD. OpenROAD reports the performance of area minimization (technology mapping) and post-routing area minimization with respect to different optimization methods. Experimental results are conducted on 104 designs from the ISCAS’85/89/99, VTR\cite{murray2020vtr}, and EPFL\cite{soeken2018epfl} benchmark suites. All experiments are conducted on an Intel Xeon Gold 6230 20x CPU. %Upon analyzing the results shown as follows, the outperformance of orchestration to the original ABC can be assured to a great extent.

\subsection{{Single Optimization Evaluations}}
\label{sec:single_result}
Initially, we validate the benefits of the orchestration concept in logic optimization for single AIG traversal. Specifically, we compiled optimization results for all 104 designs from various benchmark suites. These results are related to (1) the \textit{stand-alone optimization} methods, namely \texttt{rw}, \texttt{rs}, and \texttt{rf}, and (2) the \textit{orchestration optimization} methods, which include \textit{priority-ordered orchestration} ($\widehat{\texttt{O1}}$ – $\widehat{\texttt{O6}}$) and \textit{local-greedy orchestration} (\texttt{LocalGreedy}). A subset of these results, focusing on large designs, is presented in Table \ref{table:1O}.

The data indicates that the most effective optimization method for all designs consistently originates from one of the \textit{orchestration} methods, exhibiting notable improvements over \textit{stand-alone} optimizations. {Specifically, the best performing \textit{orchestration} algorithm ($\widehat{\texttt{O3}}$) demonstrates an average performance benefit of at least 4.0\% compared to stand-alone methods (specifically \texttt{rw}). Additionally, the table includes the average runtime cost for each optimization, where the orchestrated algorithms with better optimization performance take runtime overhead at the same time. Specifically, it takes more runtime overhead than \texttt{rs} and \texttt{rf} while less than \texttt{rw}. %The orchestration optimization requires more runtime than the stand-alone optimization, with an average runtime overhead of xx\%. 
}
%For example, for the design \texttt{voter}, the best performance produced by orchestration methods ($\widehat{\texttt{O3}}$) is 31.7\% node reduction, while the best node reduction produced by stand-alone methods (\texttt{rs}) is 20.1\%, with 11.6\% less node reduction, which confirms the advantages of implementing the orchestration concept in logic synthesis. 

\begin{figure*}
    \centering
    \begin{subfigure}[b]{0.32\textwidth}
    \centering
        \includegraphics[width=0.99\textwidth]{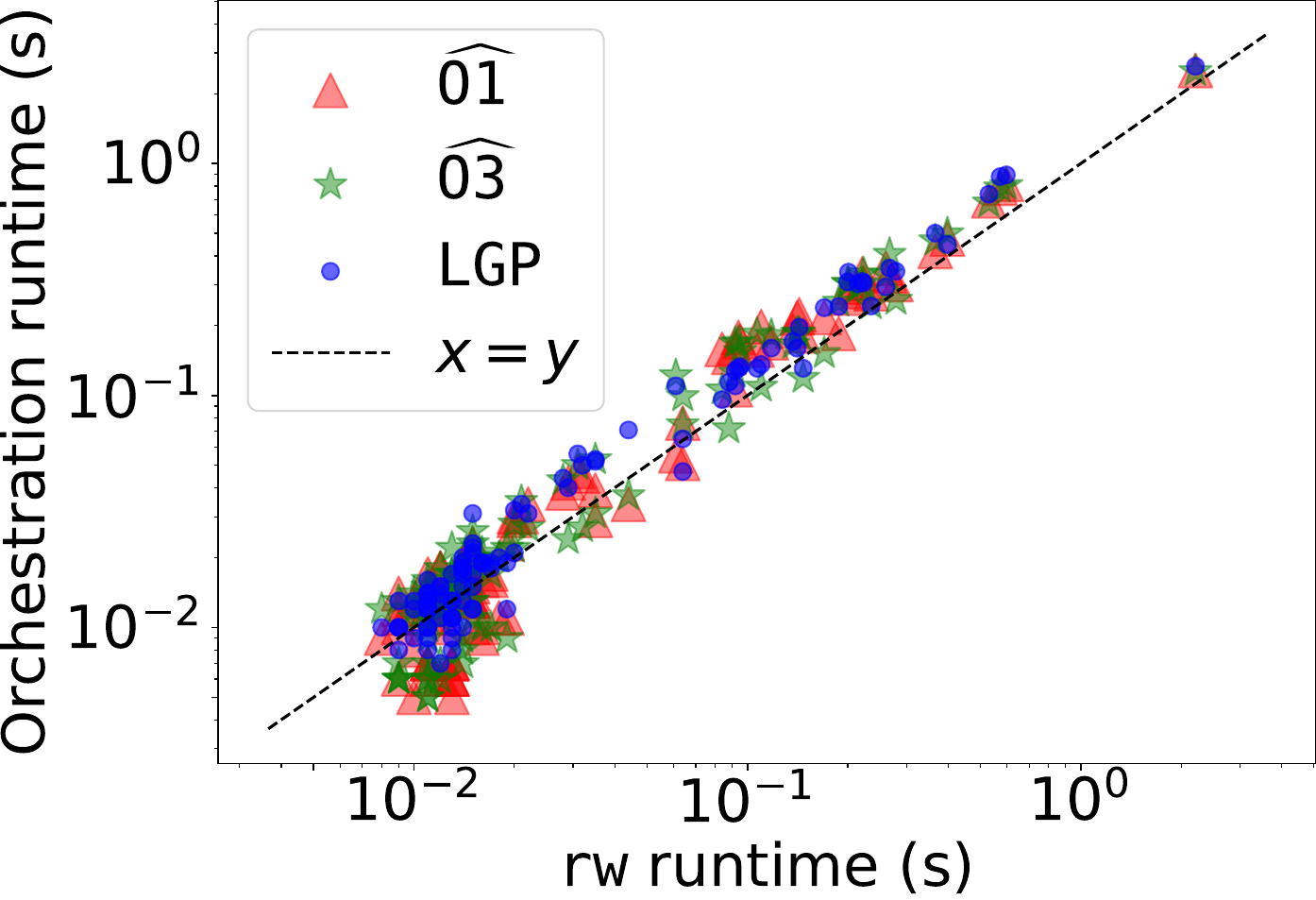}
        \caption{runtime comparison with \texttt{rw}}
        \label{fig:rt_rw}
    \end{subfigure}
    \hfill
    \begin{subfigure}[b]{0.32\textwidth}
    \centering
        \includegraphics[width=0.99\textwidth]{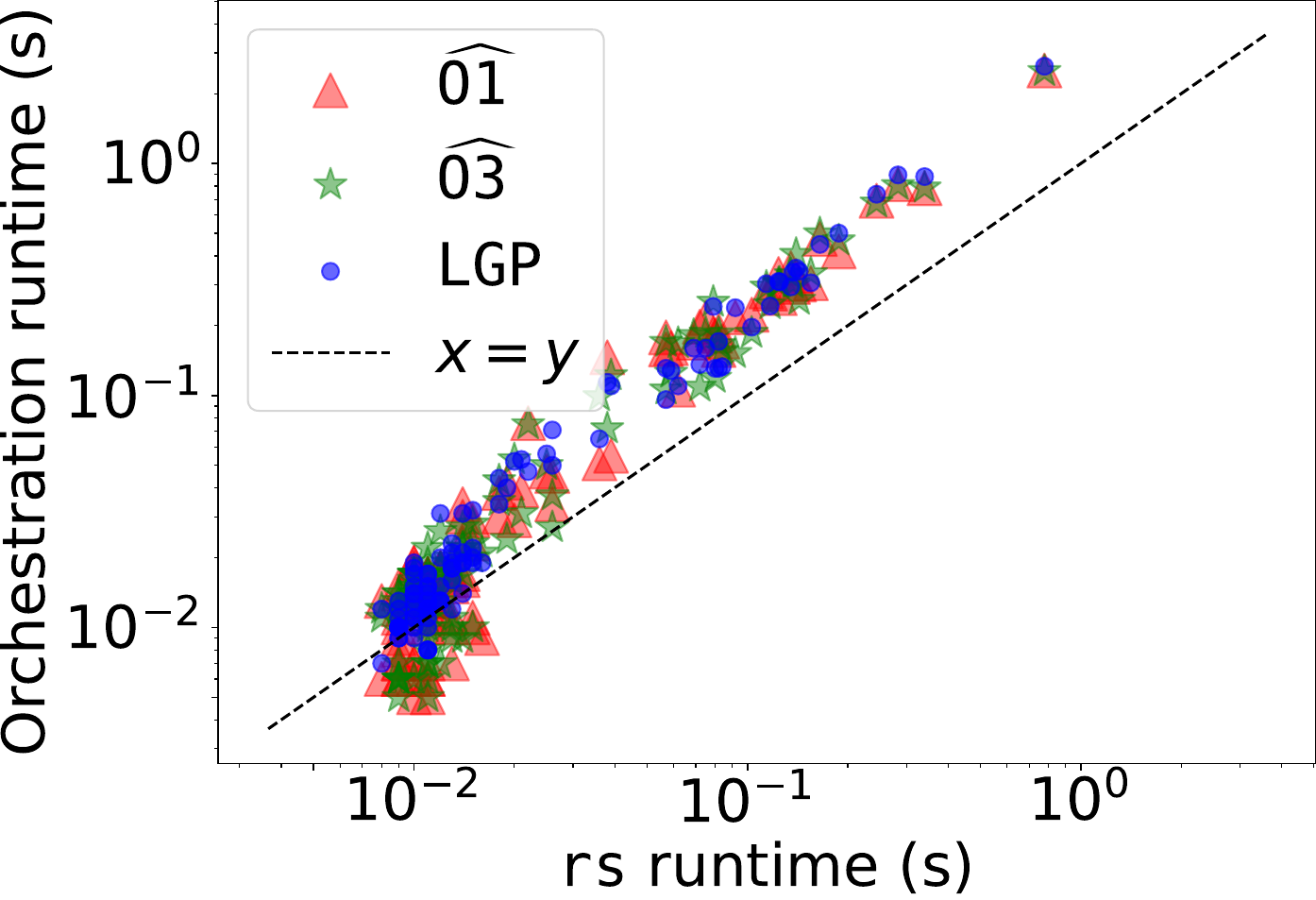}
        \caption{runtime comparison with \texttt{rs}}
        \label{fig:rt_rs}
    \end{subfigure}
    \hfill
    \begin{subfigure}[b]{0.32\textwidth}
    \centering
\includegraphics[width=0.99\textwidth]{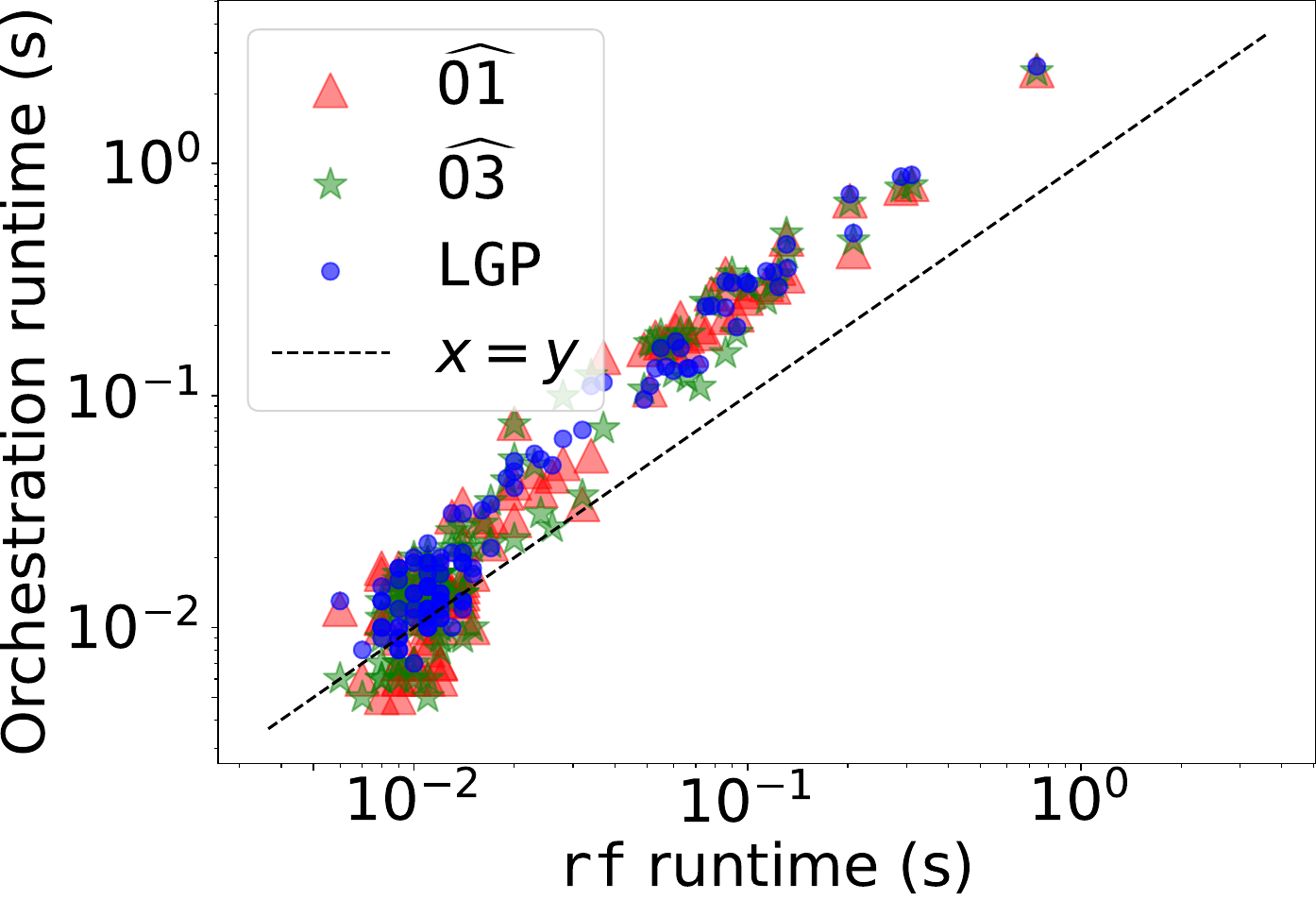}
        \caption{runtime comparison with \texttt{rf}}
        \label{fig:rt_rf}
    \end{subfigure}
     
 \caption{Runtime comparison between selected single-traversal \textit{orchestration} policies ($\widehat{\texttt{O1}}$, $\widehat{\texttt{O3}}$, and \texttt{Local-greedy} (labeled as \texttt{LGP})) and \textit{stand-alone} optimizations from ABC: (a) runtime comparison with \texttt{rw}; (b) runtime comparison with \texttt{rs}; (c) runtime comparison with \texttt{rf}.
}
    \label{fig:runtime comparison}
\end{figure*}

\subsection{Single Runtime Evaluations}
\label{sec:runtime_result}

{We also analyze the runtime of the orchestration algorithm \textit{Local-greedy} optimization (\texttt{LGP}) and \textit{Priority-ordered} optimization with $\widehat{\texttt{O1}}$ and $\widehat{\texttt{O3}}$. This analysis, including a comparison with stand-alone optimizations (i.e., \texttt{rw}, \texttt{rs}, and \texttt{rf}), is illustrated for all 104 designs in Figure \ref{fig:runtime comparison}. To effectively showcase the runtime variances, the figure employs a logarithmic scale. The x-axis represents the runtime of the stand-alone ABC optimizations, while the y-axis denotes the runtime of the \textit{orchestration} algorithms. The dotted line ($x=y$) acts as a benchmark, where points above this line indicate a higher runtime cost for the \textit{orchestration} algorithm compared to its stand-alone ABC counterpart. Conversely, points below the line suggest a lower runtime cost.} {From the runtime data, we draw two main conclusions: \textbf{(1)} Generally, orchestration algorithms {(with comparable performance, i.e., $\widehat{\texttt{O1}}$, $\widehat{\texttt{O3}}$ and \texttt{LGP})} have a comparable runtime, with \texttt{LGP} tending to incur a higher runtime overhead than other orchestration methods. \textbf{(2)} Orchestration algorithms exhibit runtime overhead when compared to stand-alone optimizations, their runtime is akin to \texttt{rw} but notably higher than \texttt{rs} and \texttt{rf}.}

{A further analysis of the runtime for each optimization iteration, focusing on different optimization methods, has been conducted. As outlined in Section \ref{sec:background_DAG_syn}, logic optimizations predominantly involve two phases: \textit{transformability check} and \textit{graph update}. \textbf{Firstly}, the transformability check constitutes the bulk of runtime in logic optimizations. \textbf{Secondly}, despite an equal number of total iterations, \texttt{rs} and \texttt{rf} optimizations are quicker than \texttt{rw}, implying that the per iteration runtime cost is lower for \texttt{rs} and \texttt{rf}. %Thus, the runtime for \texttt{rw} iterations will dominate the runtime in orchestration optimizations, resulting in comparable runtime with stand-alone \texttt{rw}. 
\textbf{Thirdly}, a substantial number of iterations are ‘wasted’ with merely performing transformability checks without contributing to graph optimization.
For instance, in Figure \ref{fig:ana}, design \textit{bfly}, the number of valid iterations is $1764$/$1374$/$920$ for \texttt{rw}/\texttt{rs}/\texttt{rf}, which is 6\%/5\%/3\% of total iterations, with 94\%/95\%/97\% iterations are wasted. However, with orchestration algorithms, the number of valid iterations is $2306$, which is 8\% of total iterations with 92\% wasted iterations. Despite the orchestration optimization has a higher percentage of valid iterations, it still incurs runtime overhead due to these wasted iterations. Specifically, in orchestration, nodes in wasted iterations undergo transformability checks for all three optimizations, significantly increasing the runtime. Particularly, \textit{Local-greedy} orchestration suffers the most as it requires transformability checks for all optimizations in every iteration. \textbf{Consequently}, the runtime inefficiency in orchestration algorithms is mainly due to the substantial number of wasted iterations involving comprehensive transformability checks. The per iteration runtime is heavily influenced by \texttt{rw} iterations, leading to an overall runtime overhead for orchestration optimizations compared to stand-alone methods, albeit being comparable to \texttt{rw}.}

\begin{table*}[t]
\centering
\caption{{Detailed results of selected large size designs. Comparison of Iterative-traversal \texttt{orchestration} with corresponding sequence optimizations from ABC.}}
\resizebox{\textwidth}{!}{%
\begin{tabular}{|c|cccccccccccccc|}
\hline
{\multirow{3}{*}{Design}} & \multicolumn{14}{c|}{AIG} \\ \cline{2-15}  
\multicolumn{1}{|c|}{} & \multicolumn{1}{c|}{Baseline} & \multicolumn{1}{c|}{\texttt{rw} $\rightarrow$ \texttt{rs} $\rightarrow$ \texttt{rf}} & \multicolumn{1}{c|}{Seq($\widehat{\texttt{O1}}$)} & \multicolumn{1}{c|}{\texttt{rw} $\rightarrow$ \texttt{rf} $\rightarrow$ \texttt{rs}} & \multicolumn{1}{c|}{Seq($\widehat{\texttt{O2}}$)} & \multicolumn{1}{c|}{\texttt{rs} $\rightarrow$ \texttt{rw} $\rightarrow$ \texttt{rf}} & \multicolumn{1}{c|}{Seq($\widehat{\texttt{O3}}$)} & \multicolumn{1}{c|}{\texttt{rs} $\rightarrow$ \texttt{rf} $\rightarrow$ \texttt{rw}} & \multicolumn{1}{c|}{Seq($\widehat{\texttt{O4}}$)} & \multicolumn{1}{c|}{\texttt{rf} $\rightarrow$ \texttt{rs} $\rightarrow$ \texttt{rw}} & \multicolumn{1}{c|}{Seq($\widehat{\texttt{O5}}$)} & \multicolumn{1}{c|}{\texttt{rf} $\rightarrow$ \texttt{rw} $\rightarrow$ \texttt{rs}} & \multicolumn{1}{c|}{Seq($\widehat{\texttt{O6}}$)} & {Seq(\texttt{LGP})} \\ \cline{2-15} 
\multicolumn{1}{|c|}{} & \multicolumn{1}{c|}{\#Node} & \multicolumn{1}{c|}{\#Node ($\Delta$\%)} & \multicolumn{1}{c|}{\#Node ($\Delta$\%)} & \multicolumn{1}{c|}{\#Node ($\Delta$\%)} & \multicolumn{1}{c|}{\#Node ($\Delta$\%)} & \multicolumn{1}{c|}{\#Node ($\Delta$\%)} & \multicolumn{1}{c|}{\#Node ($\Delta$\%)} & \multicolumn{1}{c|}{\#Node ($\Delta$\%)} & \multicolumn{1}{c|}{\#Node ($\Delta$\%)} & \multicolumn{1}{c|}{\#Node ($\Delta$\%)} & \multicolumn{1}{c|}{\#Node ($\Delta$\%)} & \multicolumn{1}{c|}{\#Node ($\Delta$\%)} & \multicolumn{1}{c|}{\#Node ($\Delta$\%)} & \#Node ($\Delta$\%) \\ \hline

\multicolumn{1}{|c|}{s38584} & \multicolumn{1}{c|}{12400} & \multicolumn{1}{c|}{10288 (17.0\%)} & \multicolumn{1}{c|}{10176 (17.9\%)} & \multicolumn{1}{c|}{10230 (17.5\%)} & \multicolumn{1}{c|}{10196 (17.8\%)} & \multicolumn{1}{c|}{10263 (17.2\%)} & \multicolumn{1}{c|}{\textbf{10115 (18.4\%)}} & \multicolumn{1}{c|}{10219 (17.6\%)} & \multicolumn{1}{c|}{10493 (15.4\%)} & \multicolumn{1}{c|}{10250 (17.3\%)} & \multicolumn{1}{c|}{10793 (13.0\%)} & \multicolumn{1}{c|}{10242 (17.4\%)} & \multicolumn{1}{c|}{10788 (13.0\%)} & 10187 (17.8\%) \\ \cline{1-15} 

\multicolumn{1}{|c|}{s35932} & \multicolumn{1}{c|}{11948} & \multicolumn{1}{c|}{8561 (28.3\%)} & \multicolumn{1}{c|}{8177 (31.6\%)} & \multicolumn{1}{c|}{8561 (28.3\%)} & \multicolumn{1}{c|}{8177 (31.6\%)} & \multicolumn{1}{c|}{8561 (28.3\%)} & \multicolumn{1}{c|}{\textbf{8177 (31.6\%)}} & \multicolumn{1}{c|}{8177 (31.6\%)} & \multicolumn{1}{c|}{8177 (31.6\%)} & \multicolumn{1}{c|}{8177 (31.6\%)} & \multicolumn{1}{c|}{8177 (31.6\%)} & \multicolumn{1}{c|}{8177 (31.6\%)} & \multicolumn{1}{c|}{\textbf{8177 (31.6\%)}} & \textbf{8129 (32.0\%)} \\ \cline{1-15} 

\multicolumn{1}{|c|}{b17\_1} & \multicolumn{1}{c|}{27647} & \multicolumn{1}{c|}{22724 (17.8\%)} & \multicolumn{1}{c|}{\textbf{22503 (18.6\%)}} & \multicolumn{1}{c|}{22664 (18.0\%)} & \multicolumn{1}{c|}{22620 (18.2\%)} & \multicolumn{1}{c|}{22831 (17.4\%)} & \multicolumn{1}{c|}{22533 (18.5\%)} & \multicolumn{1}{c|}{22841 (17.4\%)} & \multicolumn{1}{c|}{23084 (16.5\%)} & \multicolumn{1}{c|}{22888 (17.2\%)} & \multicolumn{1}{c|}{24203 (12.5\%)} & \multicolumn{1}{c|}{22832 (17.4\%)} & \multicolumn{1}{c|}{24203 (12.5\%)} & 22506 (18.6\%) \\ \cline{1-15} 

\multicolumn{1}{|c|}{b18\_1} & \multicolumn{1}{c|}{79054} & \multicolumn{1}{c|}{62622 (20.8\%)} & \multicolumn{1}{c|}{61556 (22.1\%)} & \multicolumn{1}{c|}{62425 (21.0\%)} & \multicolumn{1}{c|}{62676 (20.7\%)} & \multicolumn{1}{c|}{63060 (20.2\%)} & \multicolumn{1}{c|}{61568 (22.1\%)} & \multicolumn{1}{c|}{63042 (20.3\%)} & \multicolumn{1}{c|}{64492 (18.4\%)} & \multicolumn{1}{c|}{62552 (20.9\%)} & \multicolumn{1}{c|}{68539 (13.3\%)} & \multicolumn{1}{c|}{62288 (21.2\%)} & \multicolumn{1}{c|}{68540 (13.3\%)} & \textbf{61457 (22.3\%)} \\ \cline{1-15} 

\multicolumn{1}{|c|}{b20} & \multicolumn{1}{c|}{12219} & \multicolumn{1}{c|}{10009 (18.1\%)} & \multicolumn{1}{c|}{\textbf{9798 (19.8\%)}} & \multicolumn{1}{c|}{9972 (18.4\%)} & \multicolumn{1}{c|}{9921 (18.8\%)} & \multicolumn{1}{c|}{10029 (17.9\%)} & \multicolumn{1}{c|}{9799 (19.8\%)} & \multicolumn{1}{c|}{9910 (18.9\%)} & \multicolumn{1}{c|}{10026 (17.9\%)} & \multicolumn{1}{c|}{9933 (18.7\%)} & \multicolumn{1}{c|}{10370 (15.1\%)} & \multicolumn{1}{c|}{9907 (18.9\%)} & \multicolumn{1}{c|}{10372 (15.1\%)} & 9815 (19.7\%)\\ \cline{1-15} 

\multicolumn{1}{|c|}{b21} & \multicolumn{1}{c|}{12782} & \multicolumn{1}{c|}{10150 (20.6\%)} & \multicolumn{1}{c|}{\textbf{9904 (22.5\%)}} & \multicolumn{1}{c|}{10137 (20.7\%)} & \multicolumn{1}{c|}{10034 (21.5\%)} & \multicolumn{1}{c|}{10175 (20.4\%)} & \multicolumn{1}{c|}{9914 (22.4\%)} & \multicolumn{1}{c|}{10084 (21.1\%)} & \multicolumn{1}{c|}{10163 (20.5\%)} & \multicolumn{1}{c|}{10211 (20.1\%)} & \multicolumn{1}{c|}{10654 (16.6\%)} & \multicolumn{1}{c|}{10033 (21.5\%)} & \multicolumn{1}{c|}{10656 (16.6\%)} & 9915 (22.4\%)\\ \cline{1-15} 

\multicolumn{1}{|c|}{b22} & \multicolumn{1}{c|}{18488} & \multicolumn{1}{c|}{14960 (19.1\%)} & \multicolumn{1}{c|}{14633 (20.9\%)} & \multicolumn{1}{c|}{14891 (19.5\%)} & \multicolumn{1}{c|}{14830 (19.8\%)} & \multicolumn{1}{c|}{14952 (19.1\%)} & \multicolumn{1}{c|}{\textbf{14629 (20.9\%)}} & \multicolumn{1}{c|}{14856 (19.6\%)} & \multicolumn{1}{c|}{14962 (19.1\%)} & \multicolumn{1}{c|}{14743 (20.3\%)} & \multicolumn{1}{c|}{15654 (15.3\%)} & \multicolumn{1}{c|}{14710 (20.4\%)} & \multicolumn{1}{c|}{15656 (15.3\%)} & 14676 (20.6\%)\\ \hline

\multicolumn{1}{|c|}{bfly} & \multicolumn{1}{c|}{28910} & \multicolumn{1}{c|}{25914 (10.4\%)} & \multicolumn{1}{c|}{25750 (10.9\%)} & \multicolumn{1}{c|}{25839 (10.6\%)} & \multicolumn{1}{c|}{25945 (10.3\%)} & \multicolumn{1}{c|}{26015 (10.0\%)} & \multicolumn{1}{c|}{25818 (10.7\%)} & \multicolumn{1}{c|}{25931 (10.3\%)} & \multicolumn{1}{c|}{26163 (9.5\%)} & \multicolumn{1}{c|}{25871 (10.5\%)} & \multicolumn{1}{c|}{27373 (5.3\%)} & \multicolumn{1}{c|}{25827 (10.7\%)} & \multicolumn{1}{c|}{27375 (5.3\%)} & \textbf{25727 (11.0\%)} \\ \cline{1-15} 

\multicolumn{1}{|c|}{dscg} & \multicolumn{1}{c|}{28252} & \multicolumn{1}{c|}{25269 (10.6\%)} & \multicolumn{1}{c|}{25093 (11.2\%)} & \multicolumn{1}{c|}{25208 (10.8\%)} & \multicolumn{1}{c|}{25281 (10.5\%)} & \multicolumn{1}{c|}{25377 (10.2\%)} & \multicolumn{1}{c|}{25119 (11.1\%)} & \multicolumn{1}{c|}{25312 (10.4\%)} & \multicolumn{1}{c|}{25566 (9.5\%)} & \multicolumn{1}{c|}{25250 (10.6\%)} & \multicolumn{1}{c|}{26861 (4.9\%)} & \multicolumn{1}{c|}{25175 (10.9\%)} & \multicolumn{1}{c|}{26861 (4.9\%)} & \textbf{25052 (11.3\%)} \\ \cline{1-15} 

\multicolumn{1}{|c|}{fir} & \multicolumn{1}{c|}{27704} & \multicolumn{1}{c|}{24751 (10.7\%)} & \multicolumn{1}{c|}{24568 (11.3\%)} & \multicolumn{1}{c|}{24688 (10.9\%)} & \multicolumn{1}{c|}{24818 (10.4\%)} & \multicolumn{1}{c|}{24802 (10.5\%)} & \multicolumn{1}{c|}{24607 (11.2\%)} & \multicolumn{1}{c|}{24757 (10.6\%)} & \multicolumn{1}{c|}{24984 (9.8\%)} & \multicolumn{1}{c|}{24733 (10.7\%)} & \multicolumn{1}{c|}{26191 (5.5\%)} & \multicolumn{1}{c|}{24718 (10.8\%)} & \multicolumn{1}{c|}{26209 (5.4\%)} & \textbf{24553 (11.4\%)}\\ \cline{1-15}

\multicolumn{1}{|c|}{syn2} & \multicolumn{1}{c|}{30003} & \multicolumn{1}{c|}{26890 (10.4\%)} & \multicolumn{1}{c|}{26708 (11.0\%)} & \multicolumn{1}{c|}{26833 (10.6\%)} & \multicolumn{1}{c|}{27001 (10.0\%)} & \multicolumn{1}{c|}{26962 (10.1\%)} & \multicolumn{1}{c|}{26738 (10.9\%)} & \multicolumn{1}{c|}{26942 (10.2\%)} & \multicolumn{1}{c|}{27188 (9.4\%)} & \multicolumn{1}{c|}{26854 (10.5\%)} & \multicolumn{1}{c|}{28494 (5.0\%)} & \multicolumn{1}{c|}{26810 (10.6\%)} & \multicolumn{1}{c|}{28480 (5.1\%)} & \textbf{26700 (11.0\%)} \\ \hline

\multicolumn{1}{|c|}{div} & \multicolumn{1}{c|}{57247} & \multicolumn{1}{c|}{40965 (28.4\%)} & \multicolumn{1}{c|}{40869 (28.6\%)} & \multicolumn{1}{c|}{40965 (28.4\%)} & \multicolumn{1}{c|}{40866 (28.6\%)} & \multicolumn{1}{c|}{41006 (28.4\%)} & \multicolumn{1}{c|}{40874 (28.6\%)} & \multicolumn{1}{c|}{41004 (28.4\%)} & \multicolumn{1}{c|}{51414 (10.2\%)} & \multicolumn{1}{c|}{41142 (28.1\%)} & \multicolumn{1}{c|}{56224 (1.8\%)} & \multicolumn{1}{c|}{41104 (28.2\%)} & \multicolumn{1}{c|}{56222 (1.8\%)} & \textbf{40849 (28.6\%)}\\ \cline{1-15} 

\multicolumn{1}{|c|}{hyp} & \multicolumn{1}{c|}{214335} & \multicolumn{1}{c|}{207340 (3.3\%)} & \multicolumn{1}{c|}{206559 (3.6\%)} & \multicolumn{1}{c|}{207343 (3.3\%)} & \multicolumn{1}{c|}{211283 (1.4\%)} & \multicolumn{1}{c|}{207320 (3.3\%)} & \multicolumn{1}{c|}{\textbf{206539 (3.6\%)}} & \multicolumn{1}{c|}{206648 (3.6\%)} & \multicolumn{1}{c|}{207240 (3.3\%)} & \multicolumn{1}{c|}{206671 (3.6\%)} & \multicolumn{1}{c|}{211991 (1.1\%)} & \multicolumn{1}{c|}{206671 (3.6\%)} & \multicolumn{1}{c|}{211991 (1.1\%)} & \textbf{206530 (3.6\%)}\\ \cline{1-15} 

\multicolumn{1}{|c|}{mem\_ctrl} & \multicolumn{1}{c|}{46836} & \multicolumn{1}{c|}{46177 (1.4\%)} & \multicolumn{1}{c|}{45650 (2.5\%)} & \multicolumn{1}{c|}{46013 (1.8\%)} & \multicolumn{1}{c|}{46005 (1.8\%)} & \multicolumn{1}{c|}{46171 (1.4\%)} & \multicolumn{1}{c|}{\textbf{45360 (3.2\%)}} & \multicolumn{1}{c|}{46039 (1.7\%)} & \multicolumn{1}{c|}{45855 (2.1\%)} & \multicolumn{1}{c|}{46113 (1.5\%)} & \multicolumn{1}{c|}{46312 (1.1\%)} & \multicolumn{1}{c|}{46077 (1.6\%)} & \multicolumn{1}{c|}{46312 (1.1\%)} & 45418 (3.0\%)\\ \cline{1-15}

\multicolumn{1}{|c|}{sqrt} & \multicolumn{1}{c|}{24618} & \multicolumn{1}{c|}{19327 (21.5\%)} & \multicolumn{1}{c|}{19219 (21.9\%)} & \multicolumn{1}{c|}{19327 (21.5\%)} & \multicolumn{1}{c|}{\textbf{19218 (21.9\%)}} & \multicolumn{1}{c|}{19333 (21.5\%)} & \multicolumn{1}{c|}{19219 (21.9\%)} & \multicolumn{1}{c|}{19333 (21.5\%)} & \multicolumn{1}{c|}{19223 (21.9\%)} & \multicolumn{1}{c|}{19332 (21.5\%)} & \multicolumn{1}{c|}{23661 (3.9\%)} & \multicolumn{1}{c|}{19328 (21.5\%)} & \multicolumn{1}{c|}{23661 (3.9\%)} & \textbf{19217 (21.9\%)}\\ \cline{1-15} 

\multicolumn{1}{|c|}{voter} & \multicolumn{1}{c|}{13758} & \multicolumn{1}{c|}{8755 (36.4\%)} & \multicolumn{1}{c|}{\textbf{8428 (38.7\%)}} & \multicolumn{1}{c|}{9109 (33.8\%)} & \multicolumn{1}{c|}{10306 (25.1\%)} & \multicolumn{1}{c|}{9056 (34.2\%)} & \multicolumn{1}{c|}{8612 (37.4\%)} & \multicolumn{1}{c|}{9060 (34.1\%)} & \multicolumn{1}{c|}{8589 (37.6\%)} & \multicolumn{1}{c|}{9789 (28.8\%)} & \multicolumn{1}{c|}{12440 (9.6\%)} & \multicolumn{1}{c|}{9870 (28.3\%)} & \multicolumn{1}{c|}{12440 (9.6\%)} & 8861 (35.6\%) \\ \hline

\multicolumn{2}{|c|}{\textbf{Avg. Node Reduction\%}} & \multicolumn{1}{c|}{17.2\%} & \multicolumn{1}{c|}{\textbf{18.3\%}} & \multicolumn{1}{c|}{17.2\%} & \multicolumn{1}{c|}{16.8\%} & \multicolumn{1}{c|}{16.9\%} & \multicolumn{1}{c|}{\textbf{18.3\%}} & \multicolumn{1}{c|}{17.3\%} & \multicolumn{1}{c|}{15.8\%} & \multicolumn{1}{c|}{17.0\%} & \multicolumn{1}{c|}{9.7\%} & \multicolumn{1}{c|}{17.2\%} & \multicolumn{1}{c|}{9.7\%} & 18.2\%\\ \hline

\multicolumn{2}{|c|}{{\textbf{Avg. Runtime (s)}}} & \multicolumn{1}{c|}{{0.443}} & \multicolumn{1}{c|}{{1.162}} & \multicolumn{1}{c|}{{0.438}} & \multicolumn{1}{c|}{{1.149}} & \multicolumn{1}{c|}{{0.434}} & \multicolumn{1}{c|}{{1.148}} & \multicolumn{1}{c|}{{0.430}} & \multicolumn{1}{c|}{{1.144}} & \multicolumn{1}{c|}{{0.429}} & \multicolumn{1}{c|} {{1.143}} & \multicolumn{1}{c|} {{0.431}} & \multicolumn{1}{c|} {{1.155}} & {1.162}\\ \hline
\end{tabular}
}
% \caption{Detailed results of selected large size designs. Comparison of single-traversal \texttt{orchestration} with stand-alone optimizations from ABC.}
% \vspace{-3mm}
\label{table:3O}
\end{table*}

\begin{table*}[]
\centering
\caption{{Detailed results of selected large size designs. Comparison of \texttt{orchestration}-substituted \texttt{O-resyn}/\texttt{LGP-resyn} with original \texttt{resyn} and \texttt{resyn3}.}}
\resizebox{\textwidth}{!}{%
% \subfloat[First caption]{%
% \centering

\begin{tabular}{|cccccccccccccccc|}
\hline
\multicolumn{2}{|c|}{\multirow{3}{*}{Design}} & \multicolumn{14}{c|}{AIG: \texttt{resyn}} \\ \cline{3-16} 
\multicolumn{2}{|c|}{} & \multicolumn{2}{c|}{Baseline} & \multicolumn{2}{c|}{\texttt{resyn}} & \multicolumn{2}{c|}{\texttt{resyn3}} & \multicolumn{2}{c|}{$\widehat{\texttt{O}}$\texttt{-resyn}} & \multicolumn{2}{c|}{$\widehat{\texttt{O}}$\texttt{-resyn3}} & \multicolumn{2}{c|}{${\texttt{LGP}}$\texttt{-resyn}} & \multicolumn{2}{c|}{${\texttt{LGP}}$\texttt{-resyn3}} \\ \cline{3-16} 
\multicolumn{2}{|c|}{} & \multicolumn{1}{c|}{\#Node} & \multicolumn{1}{c|}{{Depth}} & \multicolumn{1}{c|}{\#Node ($\Delta$\%)} & \multicolumn{1}{c|}{{Depth}} & \multicolumn{1}{c|}{\#Node ($\Delta$\%)} & \multicolumn{1}{c|}{{Depth}} & \multicolumn{1}{c|}{\#Node ($\Delta$\%)} & \multicolumn{1}{c|}{{Depth}} & \multicolumn{1}{c|}{\#Node ($\Delta$\%)} & \multicolumn{1}{c|}{{Depth}} & \multicolumn{1}{c|}{\#Node ($\Delta$\%)} & \multicolumn{1}{c|}{{Depth}} & \multicolumn{1}{c|}{\#Node ($\Delta$\%)} & {Depth} \\ \hline
\multicolumn{1}{|c|}{\multirow{7}{*}{ISCAS}} & \multicolumn{1}{c|}{s38584} & \multicolumn{1}{c|}{12400} & \multicolumn{1}{c|}{36} & \multicolumn{1}{c|}{10391  (16.2\%)} & \multicolumn{1}{c|}{{25}} & \multicolumn{1}{c|}{11378  (8.2\%)} & \multicolumn{1}{c|}{28} & \multicolumn{1}{c|}{10085  (18.7\%)} & \multicolumn{1}{c|}{26} & \multicolumn{1}{c|}{{10077  (18.7\%)}} & \multicolumn{1}{c|}{{26}} & \multicolumn{1}{c|}{9988  (19.5\%)} & \multicolumn{1}{c|}{\textbf{24}} & \multicolumn{1}{c|}{\textbf{9894  (20.2\%)}} & \textbf{24} \\ \cline{2-16} 

\multicolumn{1}{|c|}{} & \multicolumn{1}{c|}{s35932} & \multicolumn{1}{c|}{11948} & \multicolumn{1}{c|}{19} & \multicolumn{1}{c|}{8518  (28.7\%)} & \multicolumn{1}{c|}{12} & \multicolumn{1}{c|}{11916  (0.3\%)} & \multicolumn{1}{c|}{19} & \multicolumn{1}{c|}{8177  (31.6\%)} & \multicolumn{1}{c|}{13} & \multicolumn{1}{c|}{8177  (31.6\%)} & \multicolumn{1}{c|}{13} & \multicolumn{1}{c|}{\textbf{8113  (32.1\%)}} & \multicolumn{1}{c|}{\textbf{11}} & \multicolumn{1}{c|}{\textbf{8113  (32.1\%)}} & \textbf{11} \\ \cline{2-16} 

\multicolumn{1}{|c|}{} & \multicolumn{1}{c|}{b17\_1} & \multicolumn{1}{c|}{27647} & \multicolumn{1}{c|}{52} & \multicolumn{1}{c|}{23021  (16.7\%)} & \multicolumn{1}{c|}{\textbf{46}} & \multicolumn{1}{c|}{26067  (5.7\%)} & \multicolumn{1}{c|}{47} & \multicolumn{1}{c|}{22046  (20.3\%)} & \multicolumn{1}{c|}{47} & \multicolumn{1}{c|}{{22011  (20.4\%)}} & \multicolumn{1}{c|}{47} & \multicolumn{1}{c|}{21475  (22.3\%)} & \multicolumn{1}{c|}{\textbf{46}} & \multicolumn{1}{c|}{\textbf{21459  (22.4\%)}} & \textbf{46} \\ \cline{2-16} 

\multicolumn{1}{|c|}{} & \multicolumn{1}{c|}{b18\_1} & \multicolumn{1}{c|}{79054} & \multicolumn{1}{c|}{132} & \multicolumn{1}{c|}{63151 (20.1\%)} & \multicolumn{1}{c|}{114} & \multicolumn{1}{c|}{70808  (10.4\%)} & \multicolumn{1}{c|}{128} & \multicolumn{1}{c|}{60231  (23.8\%)} & \multicolumn{1}{c|}{131} & \multicolumn{1}{c|}{59948  (24.2\%)} & \multicolumn{1}{c|}{130} & \multicolumn{1}{c|}{58983  (25.4\%)} & \multicolumn{1}{c|}{\textbf{127}} & \multicolumn{1}{c|}{\textbf{58710  (25.7\%)}} & \textbf{127} \\ \cline{2-16} 

\multicolumn{1}{|c|}{} & \multicolumn{1}{c|}{b20} & \multicolumn{1}{c|}{12219} & \multicolumn{1}{c|}{66} & \multicolumn{1}{c|}{10152 (16.9\%)} & \multicolumn{1}{c|}{\textbf{64}} & \multicolumn{1}{c|}{11013  (9.9\%)} & \multicolumn{1}{c|}{65} & \multicolumn{1}{c|}{9678  (20.8\%)} & \multicolumn{1}{c|}{\textbf{64}} & \multicolumn{1}{c|}{9622  (21.3\%)} & \multicolumn{1}{c|}{\textbf{64}} & \multicolumn{1}{c|}{9464  (22.5\%)} & \multicolumn{1}{c|}{65} & \multicolumn{1}{c|}{\textbf{9304  (23.9\%)}} & 65 \\ \cline{2-16} 

\multicolumn{1}{|c|}{} & \multicolumn{1}{c|}{b21} & \multicolumn{1}{c|}{12782} & \multicolumn{1}{c|}{67} & \multicolumn{1}{c|}{10211  (20.1\%)} & \multicolumn{1}{c|}{64} & \multicolumn{1}{c|}{11249  (12.0\%)} & \multicolumn{1}{c|}{65} & \multicolumn{1}{c|}{9790  (23.4\%)} & \multicolumn{1}{c|}{64} & \multicolumn{1}{c|}{9721  (23.9\%)} & \multicolumn{1}{c|}{64} & \multicolumn{1}{c|}{9580  (25.1\%)} & \multicolumn{1}{c|}{65} & \multicolumn{1}{c|}{\textbf{9414  (26.3\%)}} & \textbf{63} \\ \cline{2-16} 

\multicolumn{1}{|c|}{} & \multicolumn{1}{c|}{b22} & \multicolumn{1}{c|}{18488} & \multicolumn{1}{c|}{69} & \multicolumn{1}{c|}{15067  (18.5\%)} & \multicolumn{1}{c|}{65} & \multicolumn{1}{c|}{16643  (10.0\%)} & \multicolumn{1}{c|}{65} & \multicolumn{1}{c|}{14480  (21.7\%)} & \multicolumn{1}{c|}{65} & \multicolumn{1}{c|}{14390  (22.2\%)} & \multicolumn{1}{c|}{65} & \multicolumn{1}{c|}{14137  (23.5\%)} & \multicolumn{1}{c|}{65} & \multicolumn{1}{c|}{\textbf{13910  (24.8\%)}} & 65 \\ \hline

\multicolumn{1}{|c|}{\multirow{4}{*}{VTR}} & \multicolumn{1}{c|}{bfly} & \multicolumn{1}{c|}{28910} & \multicolumn{1}{c|}{97} & \multicolumn{1}{c|}{26177  (9.5\%)} & \multicolumn{1}{c|}{\textbf{68}} & \multicolumn{1}{c|}{26543  (8.2\%)} & \multicolumn{1}{c|}{70} & \multicolumn{1}{c|}{25242  (12.7\%)} & \multicolumn{1}{c|}{70} & \multicolumn{1}{c|}{25017  (13.5\%)} & \multicolumn{1}{c|}{70} & \multicolumn{1}{c|}{24989  (13.6\%)} & \multicolumn{1}{c|}{69} & \multicolumn{1}{c|}{\textbf{24605  (14.9\%)}} & 69 \\ \cline{2-16} 

\multicolumn{1}{|c|}{} & \multicolumn{1}{c|}{dscg} & \multicolumn{1}{c|}{28252} & \multicolumn{1}{c|}{92} & \multicolumn{1}{c|}{25427  (9.9\%)} & \multicolumn{1}{c|}{67} & \multicolumn{1}{c|}{25806  (8.7\%)} & \multicolumn{1}{c|}{68} & \multicolumn{1}{c|}{24681  (12.6\%)} & \multicolumn{1}{c|}{68} & \multicolumn{1}{c|}{24434  (13.5\%)} & \multicolumn{1}{c|}{68} & \multicolumn{1}{c|}{24274  (14.1\%)} & \multicolumn{1}{c|}{67} & \multicolumn{1}{c|}{\textbf{23945 (15.2\%)}} & \textbf{66} \\ \cline{2-16}  

\multicolumn{1}{|c|}{} & \multicolumn{1}{c|}{fir} & \multicolumn{1}{c|}{27704} & \multicolumn{1}{c|}{94} & \multicolumn{1}{c|}{24930  (10.0\%)} & \multicolumn{1}{c|}{\textbf{67}} & \multicolumn{1}{c|}{25242  (8.9\%)} & \multicolumn{1}{c|}{69} & \multicolumn{1}{c|}{24081  (13.1\%)} & \multicolumn{1}{c|}{69} & \multicolumn{1}{c|}{23870  (13.8\%)} & \multicolumn{1}{c|}{69} & \multicolumn{1}{c|}{23842  (13.9\%)} & \multicolumn{1}{c|}{\textbf{67}} & \multicolumn{1}{c|}{\textbf{23472  (15.3\%)}} & 68 \\ \cline{2-16}

\multicolumn{1}{|c|}{} & \multicolumn{1}{c|}{syn2} & \multicolumn{1}{c|}{30003} & \multicolumn{1}{c|}{93} & \multicolumn{1}{c|}{26911  (10.3\%)} & \multicolumn{1}{c|}{\textbf{67}} & \multicolumn{1}{c|}{27355  (8.8\%)} & \multicolumn{1}{c|}{68} & \multicolumn{1}{c|}{26160  (12.8\%)} & \multicolumn{1}{c|}{68} & \multicolumn{1}{c|}{25839  (13.9\%)} & \multicolumn{1}{c|}{68} & \multicolumn{1}{c|}{25806  (14.0\%)} & \multicolumn{1}{c|}{\textbf{67}} & \multicolumn{1}{c|}{\textbf{25370  (15.4\%)}} & \textbf{67} \\ \hline

\multicolumn{1}{|c|}{\multirow{5}{*}{EPFL}} & \multicolumn{1}{c|}{div} & \multicolumn{1}{c|}{57247} & \multicolumn{1}{c|}{4372} & \multicolumn{1}{c|}{40889  (28.6\%)} & \multicolumn{1}{c|}{\textbf{4359}} & \multicolumn{1}{c|}{52336  (8.6\%)} & \multicolumn{1}{c|}{4372} & \multicolumn{1}{c|}{40883  (28.6\%)} & \multicolumn{1}{c|}{4372} & \multicolumn{1}{c|}{40908  (28.5\%)} & \multicolumn{1}{c|}{4372} & \multicolumn{1}{c|}{40796  (28.7\%)} & \multicolumn{1}{c|}{4369} & \multicolumn{1}{c|}{\textbf{40749  (28.8\%)}} & 4370 \\ \cline{2-16} 

\multicolumn{1}{|c|}{} & \multicolumn{1}{c|}{hyp} & \multicolumn{1}{c|}{214335} & \multicolumn{1}{c|}{{24801}} & \multicolumn{1}{c|}{214240  (0.0\%)} & \multicolumn{1}{c|}{{24801}} & \multicolumn{1}{c|}{208371  (2.8\%)} & \multicolumn{1}{c|}{{24801}} & \multicolumn{1}{c|}{206529  (3.6\%)} & \multicolumn{1}{c|}{24801} & \multicolumn{1}{c|}{205734  (4.0\%)} & \multicolumn{1}{c|}{{24801}} & \multicolumn{1}{c|}{206005 (3.9\%)} & \multicolumn{1}{c|}{{24800}} & \multicolumn{1}{c|}{\textbf{205182  (4.3\%)}} & {\textbf{24799}} \\ \cline{2-16}

\multicolumn{1}{|c|}{} & \multicolumn{1}{c|}{mem\_ctrl} & \multicolumn{1}{c|}{46836} & \multicolumn{1}{c|}{{114}} & \multicolumn{1}{c|}{46611  (0.5\%)} & \multicolumn{1}{c|}{111} & \multicolumn{1}{c|}{46484  (0.8\%)} & \multicolumn{1}{c|}{{114}} & \multicolumn{1}{c|}{45676  (2.5\%)} & \multicolumn{1}{c|}{114} & \multicolumn{1}{c|}{45190  (3.5\%)} & \multicolumn{1}{c|}{{114}} & \multicolumn{1}{c|}{44063  (5.9\%)} & \multicolumn{1}{c|}{{111}} & \multicolumn{1}{c|}{\textbf{42165  (10.0\%)}} & \textbf{108} \\ \cline{2-16}

\multicolumn{1}{|c|}{} & \multicolumn{1}{c|}{sqrt} & \multicolumn{1}{c|}{24618} & \multicolumn{1}{c|}{{5058}} & \multicolumn{1}{c|}{19437  (21.0\%)} & \multicolumn{1}{c|}{{5058}} & \multicolumn{1}{c|}{21424  (13.0\%)} & \multicolumn{1}{c|}{{5058}} & \multicolumn{1}{c|}{{19219  (21.9\%)}} & \multicolumn{1}{c|}{{5058}} & \multicolumn{1}{c|}{{19218  (21.9\%)}} & \multicolumn{1}{c|}{{5058}} & \multicolumn{1}{c|}{\textbf{19217  (21.9\%)}} & \multicolumn{1}{c|}{{5058}} & \multicolumn{1}{c|}{\textbf{19217  (21.9\%)}} & 5058 \\ \cline{2-16} 

\multicolumn{1}{|c|}{} & \multicolumn{1}{c|}{voter} & \multicolumn{1}{c|}{13758} & \multicolumn{1}{c|}{70} & \multicolumn{1}{c|}{10446  (24.1\%)} & \multicolumn{1}{c|}{58} & \multicolumn{1}{c|}{10155  (26.2\%)} & \multicolumn{1}{c|}{68} & \multicolumn{1}{c|}{8411  (38.9\%)} & \multicolumn{1}{c|}{58} & \multicolumn{1}{c|}{8207  (40.3\%)} & \multicolumn{1}{c|}{\textbf{57}} & \multicolumn{1}{c|}{8224  (40.2\%)} & \multicolumn{1}{c|}{\textbf{57}} & \multicolumn{1}{c|}{\textbf{8071  (41.3\%)}} & 58 \\ \hline

\multicolumn{4}{|c|}{\textbf{Avg. Node Reduction\%}} & \multicolumn{2}{c|}{15.7\%} & \multicolumn{2}{c|}{8.9\%} & \multicolumn{2}{c|}{19.2\% (+3.5\%)} & \multicolumn{2}{c|}{19.7\% (+10.8\%)} & \multicolumn{2}{c|}{20.4\% (+4.7\%)} & \multicolumn{2}{c|}{\textbf{21.4\% (+11.5\%)}} \\ \hline

\multicolumn{4}{|c|}{{\textbf{Avg. Runtime (s)}}} & \multicolumn{2}{c|}{{0.717}} & \multicolumn{2}{c|}{{0.521}} & \multicolumn{2}{c|}{{1.148}} & \multicolumn{2}{c|}{{1.840}} & \multicolumn{2}{c|}{{1.197}} & \multicolumn{2}{c|} {{1.908}} \\ \hline
\end{tabular}%
% }
}
% \caption{Detailed results of selected large size designs. Comparison of \texttt{orchestration}-substituted \texttt{O-resyn} with original \texttt{resyn}/\texttt{resyn3}.}
%\vspace{-4mm}
\label{table:resyn}
\end{table*}

\begin{table*}[]
\caption{The results reported by OpenROAD with orchestration methods implementation, including the results from logic synthesis, i.e., AIG minimization, technology mapping with nangate 45nm, and post-routing.}
\resizebox{\textwidth}{!}{%
\begin{tabular}{|c|c|c|c|c|c|c|c|c|c|}
\hline
 &    &   &   & &  &  &  &  &      \\
\multirow{-2}{*}{} & \multirow{-2}{*}{\begin{tabular}[c]{@{}c@{}}Logic Synthesis \\ (resyn)\end{tabular}} & \multirow{-2}{*}{\begin{tabular}[c]{@{}c@{}}Logic Synthesis\\ (LGP-resyn)\end{tabular}} & \multirow{-2}{*}{\begin{tabular}[c]{@{}c@{}}Logic Synthesis\\  (O-resyn)\end{tabular}} & \multirow{-2}{*}{\begin{tabular}[c]{@{}c@{}}Tech Map\\  (resyn)\end{tabular}} & \multirow{-2}{*}{\begin{tabular}[c]{@{}c@{}}Tech Map\\ (LGP-resyn)\end{tabular}} & \multirow{-2}{*}{\begin{tabular}[c]{@{}c@{}}Tech Map\\  (O-resyn)\end{tabular}} & \multirow{-2}{*}{\begin{tabular}[c]{@{}c@{}}Post-routing \\ (resyn)\end{tabular}} & \multirow{-2}{*}{\begin{tabular}[c]{@{}c@{}}Post-routing\\ (LGP-resyn)\end{tabular}} & \multirow{-2}{*}{\begin{tabular}[c]{@{}c@{}}Post-routing\\ (O-resyn)\end{tabular}} \\ \hline
   & Node   & Node   & Node  & Area      & Area & Area    & Area/$um^{2}$       & Area/$um^{2}$       & Area/$um^{2}$        \\ \hline
s38584             & 10391                                                                                & 9988 (-3.9\%)                                                                             & 10085 (-2.9\%)                                                                                & 13161.95                                                                      & 13000.48 (-1.2\%)                                                                   & 13011.922 (-1.1\%)                                                                     & 14313                                                                             & 14013 (-2.1\%)                                                                          & 14137 (-1.2\%)                                                                            \\ \hline
s35932             & 8518   & 8113 (-4.8\%)    & 8177 (-4.0\%)                                                         & 15368.15                                              & 15372.40 (+0.03\%)              &15368.15 (0)   & 16045      & 16055 (+0.06\%)        & 16045 (0)                                                         \\ \hline
b17\_1             & 23021                                                                                & 21475 (-6.7\%)                                                                             & 22046 (-4.2\%)                                                                                & 26798.44                                                                      & 25640.27 (-4.3\%)                                                                                    & 26019.588 (-2.9\%)                                                                     & 29138                                                                             &   28193 (-3.2\%)                                                                                      & 28561 (-2.0\%)                                                                            \\ \hline
b18\_1             & 63151                                                                                & 58983 (-6.6\%)                                                                             & 60231 (-4.6\%)                                                                                & 70259.38                                                                      & 68211.18 (-2.9\%)                                                                   & 67852.876 (-3.4\%)                                                                     & 76811                                                                             & 74499 (-3.0\%)                                                                          & 74516 (-3.0\%)                                                                            \\ \hline
b20                & 10152                                                                                & 9464 (-6.8\%)                                                                              & 9678 (-4.6\%)                                                                                 & 11295.96                                                                      & 10915.57 (-3.3\%)                                                                   & 11029.158 (-2.4\%)                                                                     & 12247                                                                             & 11861 (-3.1\%)                                                                          & 12027 (-1.8\%)                                                                            \\ \hline
b21                & 10211                                                                                & 9580 (-6.2\%)                                                                              & 9790 (-4.1\%)                                                                                 & 11585.63                                                                      & 11190.09 (-3.4\%)                                                                   & 11222.274 (-3.1\%)                                                                     & 12596                                                                             & 12246 (-2.7\%)                                                                          & 12276 (-2.5\%)                                                                            \\ \hline
b22                & 15067                                                                                & 14137 (-6.2\%)                                                                             & 14480 (-3.9\%)                                                                                & 16115.61                                                                      & 15879.67 (-1.4\%)                                                                   & 15928.612 (-1.2\%)                                                                     & 17856                                                                             & 17404 (-2.5\%)                                                                          & 17518 (-1.9\%)                                                                            \\ \hline
\end{tabular}
}
% \caption{The results reported by OpenROAD with orchestration methods implementation, including the results from logic synthesis, i.e., AIG minimization, technology mapping with nangate 45nm, and post-routing.}
\label{tbl:resyn_openroad}
\end{table*}

\subsection{Iterative Optimization Evaluations}
\label{sec:iterative_result}

It is known that DAG-aware synthesis performs better in iterative transformations. However, considering the runtime for fair comparison, in this iterative optimization evaluation, %we omit the iterative optimization with \textit{Local-greedy} orchestration policy due to its significant runtime overhead even with single traversal optimization as shown in Section \ref{sec:runtime_result}. 
%Therefore, to make fair comparison, 
we compare \textit{priority-ordered orchestration} optimization (e.g., \{$\widehat{\texttt{O1}}$ $\rightarrow$ $\widehat{\texttt{O1}}$ $\rightarrow$ $\widehat{\texttt{O1}}$\}, denoted as Seq($\widehat{\texttt{O1}}$)) to the corresponding stand-alone optimization sequence of the priority order (e.g., correspond to $\widehat{\texttt{O1}}$, the sequence is \{\texttt{rw} $\rightarrow$ \texttt{rs} $\rightarrow$ \texttt{rf}\}, denoted as Seq(ABC)). We use the same notations and perform experiments on other \textit{Priority-ordered} orchestration algorithms. 
{The results of the iterative-traversal with orchestration algorithms and the corresponding sequence of stand-alone optimizations are shown in Table \ref{table:3O}. In all permutations of stand-alone optimization sequences, node reduction performance ranges from 16.9\% to 17.2\%. However, with orchestrated operation sequences, this performance varies between 9.7\% and 18.3\%. In line with single traversal results, sequential optimizations using $\widehat{\texttt{O1}}$ and $\widehat{\texttt{O3}}$ surpass their corresponding stand-alone sequences by 1.1\% and 1.4\% respectively.}
%Based on results shown in Section \ref{sec:single_result} that demonstrates $\widehat{\texttt{O1}}$ and $\widehat{\texttt{O3}}$ are the best among \texttt{Priority} {orchestration} policies, we keep only these two priority order policies for experiments beyond this point. 

%Figure \ref{fig:3O} includes the results from the experiments mentioned above. Note that we only crop range $\left [0, 0.20\right]$, since iterative optimization significantly reduces the AIG size, to show the most clustered part. {We set \textbf{R} as the ratio of the number of designs on which Seq($\widehat{\texttt{O1}}$) and Seq($\widehat{\texttt{O3}}$) outperform Seq(ABC) to the number of total tested designs. Overall, both Seq($\widehat{\texttt{O1}}$) and Seq($\widehat{\texttt{O3}}$) outperform corresponded Seq(ABC) on more than $93\%$ of the benchmarks.} 

{Furthermore, we evaluate the performance of the orchestration methods when combined with other orthogonal optimizations in a sequential synthesis flow. Specifically, we evaluate the \textit{orchestration} algorithm in \texttt{resyn} and \texttt{resyn3} in ABC. The original flow involves iterative transformations such as \textit{rewriting} (\texttt{rw}), \textit{resubstitution} (\texttt{rs}), \textit{refactoring} (\texttt{rf}), and \textit{balance} (\texttt{b}). The zero-cost replacement enabled \texttt{rw}, \texttt{rs}, and \texttt{rf} are denoted as \texttt{rwz}, \texttt{rsz}, and \texttt{rfz}, respectively. Similarly for the zero-cost replacement enabled orchestration algorithms are denoted as \texttt{Z1} to \texttt{Z6}.
The optimization flow in \texttt{resyn} is \{\texttt{b};\texttt{rw};\texttt{rwz};\texttt{b};\texttt{rwz};\texttt{b}\}; the flow of \texttt{resyn3} is \{\texttt{b};\texttt{rs};\texttt{rs -K 6};\texttt{b};\texttt{rsz};\texttt{rsz -K 6};\texttt{b};\texttt{rsz -K 5};\texttt{b}\}.
We follow the permutation of the original flows by replacing the stand-alone optimization with orchestration optimizations to compose {orchestration} flows. 
We name the resyn flow where \texttt{rw}/\texttt{rwz} is replaced with $\widehat{\texttt{O1}}$/$\widehat{\texttt{Z1}}$ (because \texttt{rw} has the highest priority in $\widehat{\texttt{O1}}$) as \texttt{O-resyn}, the resyn3 flow where \texttt{rs} is replaced with $\widehat{\texttt{O3}}$ (because \texttt{rs} has the highest priority in $\widehat{\texttt{O3}}$) as \texttt{O-resyn3}, and the resyn/resyn3 flow where \texttt{rw}(\texttt{rwz})/\texttt{rs} is replaced with \texttt{Local-greedy}(\texttt{Local-greedy-z}) as \texttt{LGP-resyn}/\texttt{LGP-resyn3}. 

Table \ref{table:resyn} shows the AIG optimization results of 16 designs. Upon comparing the average node reduction of each optimization option, we can observe a consistent improvement with orchestration synthesis flows. Specifically, ${\texttt{O}}$\texttt{-resyn} and ${\texttt{LGP}}$\texttt{-resyn} outperforms \texttt{resyn} by 3.5\% and 4.7\% more average node reductions, respectively; and ${\texttt{O}}$\texttt{-resyn3} and ${\texttt{LGP}}$\texttt{-resyn3} with 10.8\% and 11.5\% more average node reductions than \texttt{resyn3}.}

%We conduct the logic synthesis with three resyn flow on all 104 design and present the results for 7 designs in the first three columns in Table \ref{tbl:resyn_openroad}. Both orchestration flows, i.e., \texttt{O-resyn} and \texttt{LGP-resyn}, consistently outperform the original \texttt{resyn} for all designs. For most designs (6/7), the \texttt{O-resyn} can slightly outperform \texttt{LGP-resyn}. The observations further confirm the advantages of implementing orchestration in logic synthesis for iterative optimizations and combined optimizations with other optimization methods.

%\input{tbls/table_resyn_map}

%\vspace{-3mm}
\subsection{End-to-end Evaluations}
\label{sec:eval_openroad}
\begin{figure}
    \centering
    \includegraphics[width=0.6\linewidth]{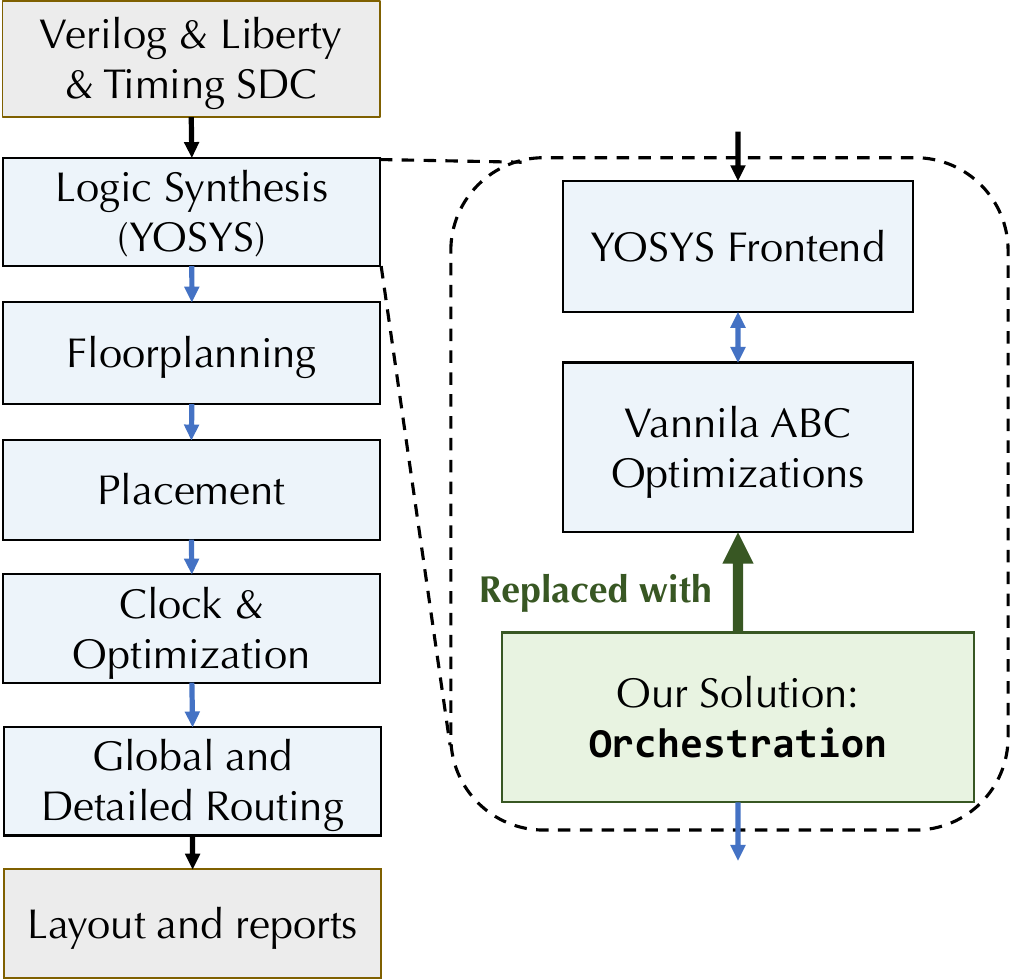}
    \caption{The OpenROAD framework integrated with proposed orchestration methods. The dash line blue box shows the details in logic synthesis where the original ABC is replaced with our proposed orchestration optimization implemented ABC.  }
    \label{fig:openroad_frame}
\end{figure}

Finally, we integrate our proposed orchestration optimization methods into the end-to-end design framework OpenROAD (Open Resilient Design for Autonomous Systems) \cite{ajayi2019openroad} to evaluate the end-to-end performance by the orchestration improved logic synthesis.
{OpenROAD \cite{ajayi2019openroad} Project is an open-source project aiming at developing a comprehensive, end-to-end, automated IC (Integrated Circuit) design flow that supports a wide range of design styles and technology nodes. It integrates various open-source tools to streamline chip development. The flow begins with RTL synthesis, where Yosys \cite{wolf2016yosys} converts high-level RTL descriptions into gate-level netlists and performs logic synthesis and technology mapping via ABC \cite{mishchenko2007abc}. As shown in Figure \ref{fig:openroad_frame}, this is the specific integration where we deploy our proposed orchestration methods in ABC in the end-to-end design flow (the dash line box). Next, the OpenROAD flow performs floorplanning, placement, and global routing. Tools such as RePlAce, TritonRoute, and FastRoute are used for these tasks, respectively. Afterward, detailed routing and signoff checks are completed, using tools like OpenROAD's built-in router and Magic.}

%To make fair comparisons, we follow the permutation of the original flows by replacing \texttt{rw} with $\widehat{\texttt{O1}}$ or \texttt{LocalGreedy} to compose \texttt{orchestration} flows, namely \texttt{O-resyn}. For instance, $\widehat{\texttt{O1}}$\texttt{-resyn} comprises \{\texttt{balance} $\rightarrow$ $\widehat{\texttt{O1}}$ $\rightarrow$ $\widehat{\texttt{O1}}$ $\rightarrow$ \texttt{balance} $\rightarrow$ $\widehat{\texttt{O1}}$ $\rightarrow$ \texttt{balance}\}. Table \ref{table:resyn} shows the AIG optimization results of 16 designs. Upon comparing the average node reduction of each optimization option, we can observe a consistent improvement with \texttt{O-resyn}. Specifically, $\widehat{\texttt{O1}}$\texttt{-resyn} and $\widehat{\texttt{O3}}$\texttt{-resyn} outperforms \texttt{resyn} by 4.4\% and 4.0\% more average node reductions, respectively; and $\widehat{\texttt{O1}}$\texttt{-resyn3} and $\widehat{\texttt{O3}}$\texttt{-resyn3} with 10.7\% and 10.8\% more average node reductions than \texttt{resyn3}. 

\subsubsection{Technology Mapping}
\label{sec:eval_map}

We have implemented AIG technology mapping for standard cells using the 45nm Nangate library \cite{knudsen2008nangate} and applied \texttt{resyn}, \texttt{O-resyn}, and \texttt{LGP-resyn} across all 104 designs in a consistent environment. Selected results for 7 detailed cases are presented in Table \ref{tbl:resyn_openroad} (columns 5 – 7), with the technology mapping outcomes reported by Yosys in OpenROAD. Generally, flows incorporating orchestration optimizations tend to yield better area minimization, averaging 2.2\% more area reduction. This suggests the potential of integrating \textit{orchestration} into existing synthesis flows for enhanced technology mapping performance. However, an exception is observed in the case of \textit{s35932}, where although orchestration-enhanced \texttt{resyn} flows surpass the original \texttt{resyn} in AIG reduction, they result in larger areas post-technology mapping.

Furthermore, a comparison between the post-technology mapping results and those from logic synthesis reveals that the benefits gained from orchestration methods during logic synthesis can diminish, disappear, or even turn into drawbacks after technology mapping. This discrepancy likely arises from the misalignment between technology-independent logic synthesis and technology-dependent mapping cost models, attributable to the high-level abstractions involved at the logic level.%, which introduces us to go further in exploring the miscorrelations in the successive stages. 

\subsubsection{Post-Routing}

Furthermore, we carry out post-routing evaluations in OpenROAD, applying the three \texttt{resyn} flows to various designs. The results, detailed in the last three columns of Table \ref{tbl:resyn_openroad}, indicate that the orchestration-enhanced flows (\texttt{O-resyn} and \texttt{LGP-resyn}) generally maintain superiority over the original \texttt{resyn} across most designs. However, the margin of this superiority is reduced when compared to the gains observed in logic synthesis. For instance, in the case of the design \textit{b21}, the \texttt{LGP-resyn} flow demonstrates a 6.2\% improvement in AIG reduction, but this advantage is reduced to 2.7\% in terms of area minimization following post-routing. A notable exception is observed in the design \textit{s35932}, where, despite a 4.8\% improvement in AIG reduction with orchestration methods, the post-routing area minimization performance degrades. This trend, similar to what was observed in technology mapping, underscores the potential misalignments between the benefits achieved during technology-independent logic synthesis and the outcomes post technology-dependent mapping and routing stages. 
%In conclusion, although the performance after post-routing correlates poorly with the performance after logic synthesis, it correlates well with that after technology mapping, which motivates us for implementing technology-aware logic synthesis in the futher works. 

In conclusion, our study reveals a modest correlation between the improvements achieved in logic optimization and the enhancements in post-routing performance. However, there is a more pronounced connection between the results following technology mapping and those observed in the post-routing stage. This finding motivates the focus of our future research on developing technology-aware logic synthesis approaches, aiming to align more closely with the subsequent stages of technology mapping and routing, thereby enhancing the overall design efficiency.
\section{Conclusion}
\label{sec:conclusion}

In this work, we propose a novel concept in logic synthesis development -- DAG-aware synthesis \textit{orchestration}, which encompasses multiple optimization operations within a single AIG traversal. 
The proposed concept is implemented in ABC, orchestrating the pre-exisiting stand-alone optimizations, namely \textit{rewriting}, \textit{resubstitution}, \textit{refactoring} for fine-grained node-level logic optimization within a single AIG traversal. Specifically, we provide two algorithms for this \textit{orchestration} process: (1) The \textit{Local-greedy orchestration} algorithm, which selects the optimization operation offering the highest local gain at each node for AIG optimization; (2) The \textit{Priority-ordered orchestration} algorithm, which employs a predefined priority order to select the optimization operation at each node.
Our implementations have been rigorously tested on 104 designs from benchmark suites such as ISCA'85/89/99, VTR, and EPFL. {In comparison to conventional stand-alone optimizations, our \textit{orchestration} optimization achieves superior performance with a reasonable runtime overhead during single graph traversal. Additionally, this optimization maintains its performance benefits in iterative optimizations and integrated design flows, such as \texttt{resyn}, when combined with other optimizations like \textit{balance}. Notably, when implemented within an end-to-end design flow, the orchestration algorithm surpasses stand-alone optimizations in technology mapping and post-routing for the majority of designs. However, it is important to note the observed discrepancies between technology-independent stages (e.g., logic synthesis) and technology-dependent stages (e.g., technology mapping and post-routing). These observations have spurred our interest in future research, specifically aiming to develop end-to-end aware DAG-aware synthesis \textit{orchestrations} that address these optimization miscorrelations.}

\newpage 
\bibliographystyle{IEEEtran}
\bibliography{main}

%\begin{biography}[{\includegraphics[width=1in,height=1.25in,clip,keepaspectratio]{mshell}}]{Michael Shell}

\begin{IEEEbiography}[{\includegraphics[width=1in,height=1.25in,clip,keepaspectratio]{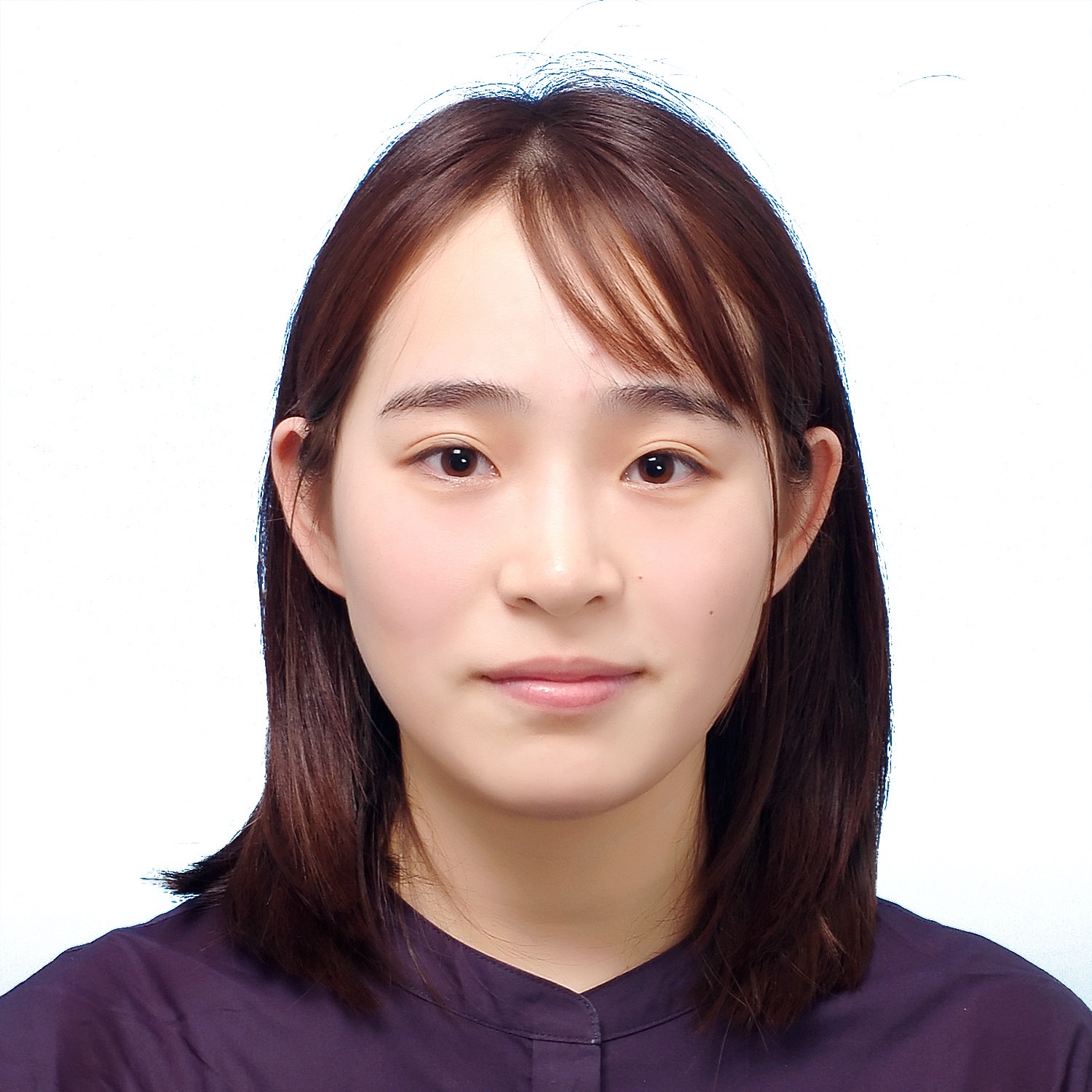}}]{Yingjie Li} (Student Member, IEEE) is currently a fourth-year PhD candidate in Computer Engineering at the University of Maryland, College Park under the supervision of Prof. Cunxi Yu. Her research focuses on physics-aware infrastructure for optical computing platforms, hardware-software co-design, and efficient AI/ML algorithms. She also works in electronic design automation (EDA), focusing on machine learning for synthesis and verification. She received B.S. degree in 2018 from Huazhong University of Science and Technology in Wuhan, China, and her master degree from Cornell University in 2019. Her work received the Best Paper Award at DAC (2023), American Physical Society DLS poster award (2022) and Best Poster Presentation Award at DAC Young Fellow (2020). Yingjie won the Second Place at the ACM/SIGDA Student Research Competition (2023) and was selected as the EECS Rising Star (2023).

\end{IEEEbiography}

\begin{IEEEbiography}[{\includegraphics[width=1in,height=1.25in,clip,keepaspectratio]{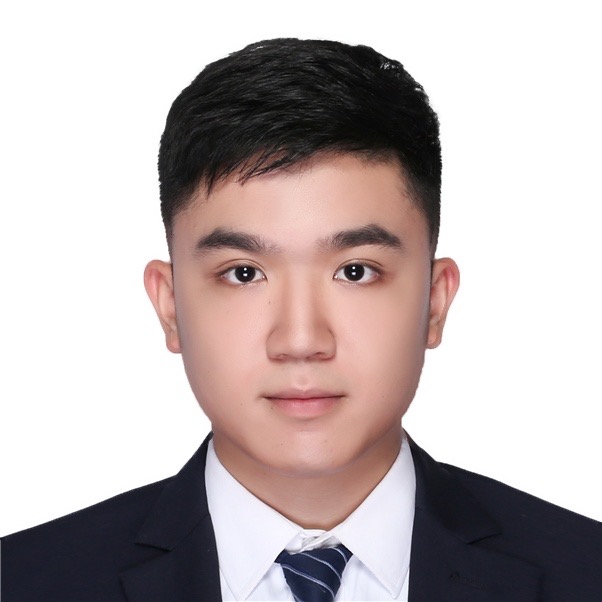}}]{Mingju Liu} (Student Member, IEEE) is currently a second-year PhD student in Computer Engineering at the University of Maryland, College Park under the supervision of Prof. Cunxi Yu. His research spans across hardware-software co-design of deep learning algorithms, focusing on Electronic Design Automation (EDA) challenges. He also works on projects in logic synthesis and combinatorial optimization. He received his B.S. degree in 2020 from the University of Electronic Science and Technology of China in Chengdu, China, and his master's degree from Rutgers University in 2021. He was selected into the DAC Young Fellows Program in 2023.

\end{IEEEbiography}

\begin{IEEEbiography}[{\includegraphics[width=1in,height=1.25in,clip,keepaspectratio]{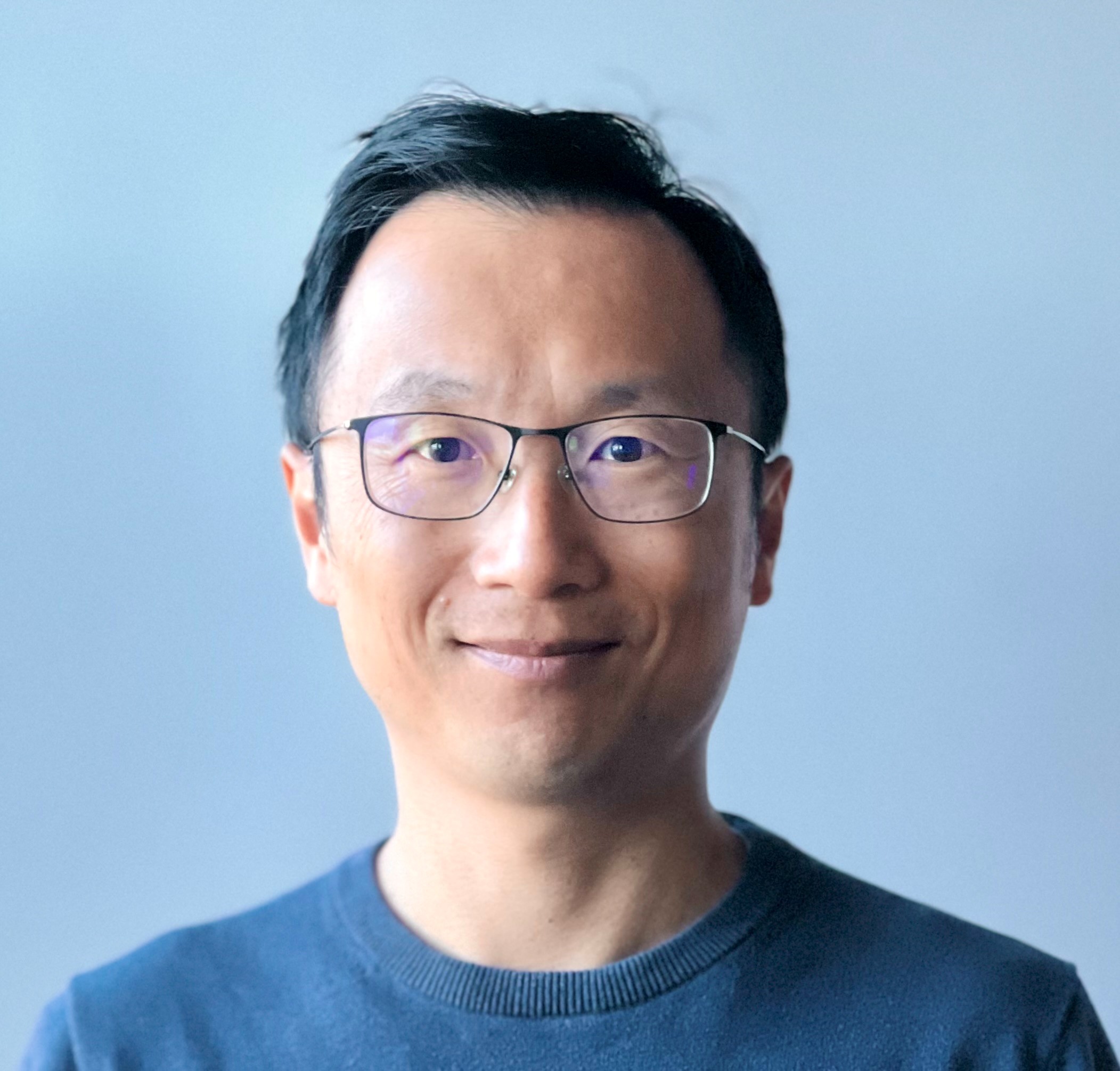}}]{Haoxing Ren} (Fellow, IEEE) is the Director of Design Automation Research at NVIDIA, focusing on leveraging machine learning and GPU-accelerated tools to enhance chip design quality and productivity. Before joining NVIDIA in 2016, he dedicated 15 years to EDA algorithm research and design methodology innovation at IBM Microelectronics and IBM Research. Haoxing is widely recognized for his contributions to physical design, AI, and GPU acceleration for EDA, which have earned him several prestigious awards, including the IBM Corporate Award and best paper awards at ISPD, DAC, TCAD, and MLCAD. He holds over twenty patents and has co-authored over 100 papers and books, including a book on ML for EDA and several book chapters in physical design and logic synthesis. He holds Bachelor's and Master's degrees from Shanghai Jiao Tong University and Rensselaer Polytechnic Institute, respectively, and earned his Ph.D. from the University of Texas at Austin. He is a Fellow of the IEEE.
\end{IEEEbiography}

\begin{IEEEbiography}[{\includegraphics[width=1in,height=1.25in,clip,keepaspectratio]{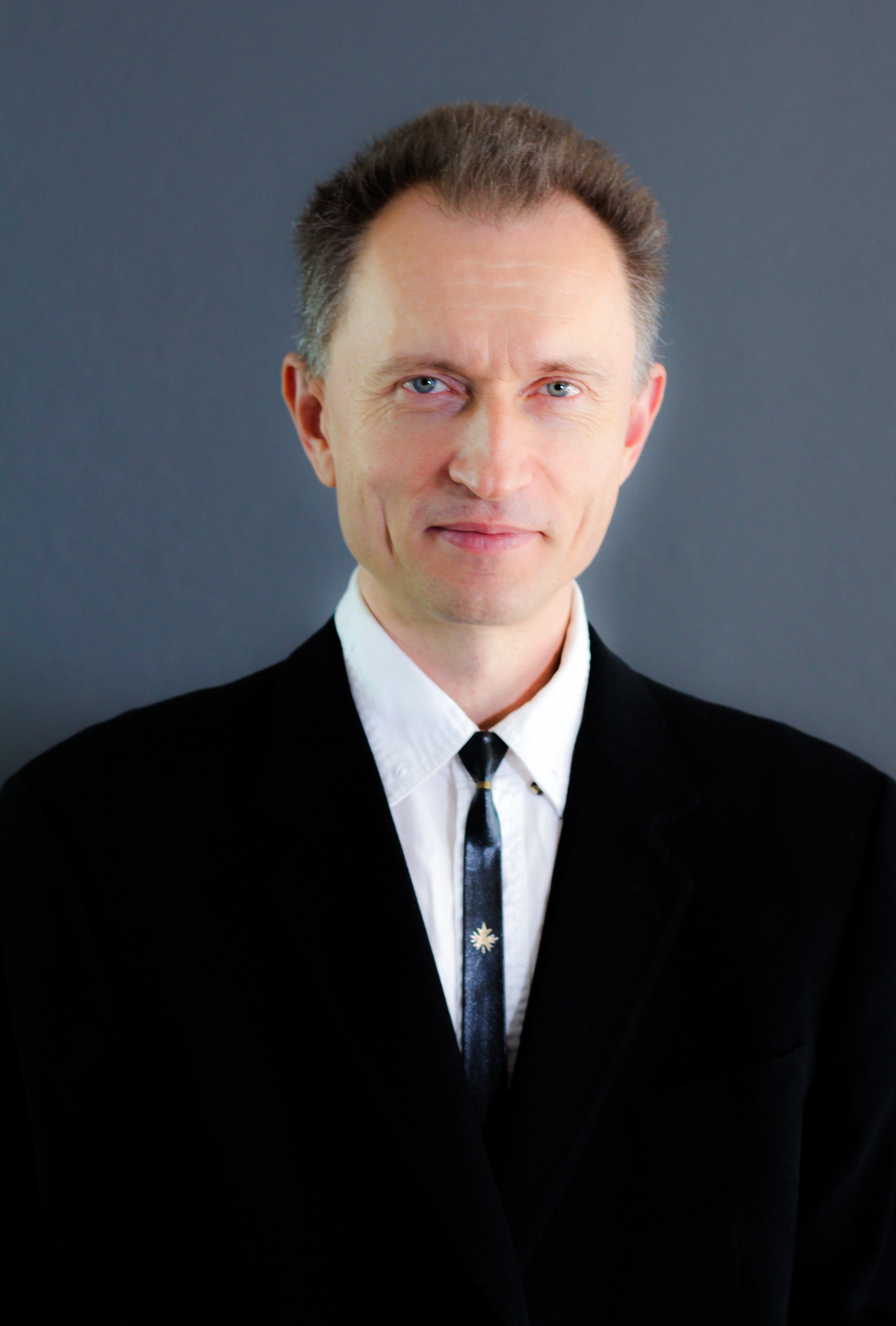}}]{Alan Mishchenko} (Senior Member, IEEE)received the M.S. degree from the Moscow Institute of Physics and Technology, Moscow, Russia, in 1993 and the Ph.D. degree from the Glushkov Institute of Cybernetics, Kiev, Ukraine, in 1997. In 2002, he joined the EECS Department, University of California at Berkeley, Berkeley, CA, USA, where he is currently a Full Researcher. His current research interests include computationally efficient logic synthesis, formal verification, and machine learning.
\end{IEEEbiography}

\begin{IEEEbiography}[{\includegraphics[width=1in,height=1.25in,clip,keepaspectratio]{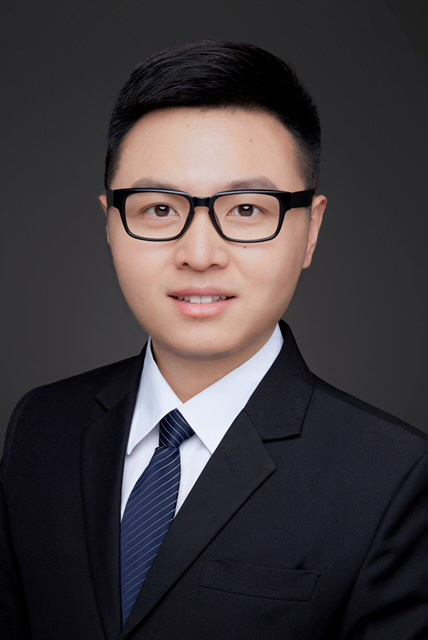}}]{Cunxi Yu} (Member, IEEE)
is an Assistant Professor in the ECE Department at the University of Maryland, College Park. His research interests focus on novel algorithms, systems, and hardware designs for computing and security. Before joining University of Maryland, Cunxi was an Assistant Professor at University of Utah, PostDoc at Cornell University in 2018-2019 and EPFL in 2017-2018, and was a research intern at IBM T.J Watson Research Center (2015, 2016). He received Ph.D. degree from UMass Amherst in 2017. His work received the best paper nomination at ASP-DAC (2017), TCAD Best paper nomination (2018), 1st place at DAC Security Contest (2017), NSF CAREER Award (2021), DLS Best Poster Honorable Mention at (2022), and Best Paper Award at DAC (2023). He served as Organizing Committee in IWLS, ICCD, VLSI-SoC, ASAP, as a TPC member in ICCAD, DATE, ASP-DAC, DAC, and General Chair of IWLS 2023.
\end{IEEEbiography}

%% insert where needed to balance the two columns on the last page with
%%\newpage

% ==== SWITCH OFF the BIO for submission
% ==== SWITCH OFF the BIO for submission


\vfill

\end{document}